\documentclass[flushbottom, nobalancelastpage,superscriptaddress,aps,prd,showpacs,nofootinbib,floatfix,twocolumn]{revtex4-1}

\usepackage[english]{babel}
\usepackage{graphicx}
\usepackage{amsmath}
\usepackage{amssymb}
\usepackage[subrefformat=parens,labelformat=parens]{subfig}
\usepackage{units}
\usepackage{xspace}
\usepackage{booktabs}
\usepackage[switch]{lineno}

\def\EeV{\ifmmode {\mathrm{Ee\kern -0.07em V}}\else
                   \textrm{Ee\kern -0.07em V}\fi}
\def\TeV{\ifmmode {\mathrm{Te\kern -0.07em V}}\else
                   \textrm{Te\kern -0.07em V}\fi}
\def\eV{\ifmmode {\mathrm{\ e\kern -0.07em V}}\else
                   \textrm{e\kern -0.07em V}\fi}
\def\gcm{\ensuremath{\mathrm{g/cm}^2}\xspace}
\def\Xmax{\ensuremath{X_\mathrm{max}}\xspace}
\def\sigmaXmax{\ensuremath{\sigma(X_\text{max})}\xspace}
\def\meanXmax{\ensuremath{\langle X_\text{max}\rangle}\xspace}
\def\meanLnA{\ensuremath{\langle\ln A\rangle}\xspace}

\newcommand{\gcmdec}{\ifmmode {\gcm/\mathrm{decade}}\else
                     {\gcm/decade}\fi\xspace}%

\newcommand{\energy}[1]{\ensuremath{10^{#1}}\,\eV}

\newcommand{\bea}{\begin{linenomath}\begin{eqnarray}}
\newcommand{\eea}{\end{eqnarray}\end{linenomath}}
\newcommand{\bean}{\begin{linenomath}\begin{eqnarray*}}
\newcommand{\eean}{\end{eqnarray*}\end{linenomath}}
\newcommand{\be}{\begin{linenomath}\begin{equation}}
\newcommand{\ee}{\end{equation}\end{linenomath}}
\newcommand{\ben}{\begin{linenomath}\begin{equation*}}
\newcommand{\een}{\end{equation*}\end{linenomath}}

\newcommand{\bem}{\begin{linenomath}\begin{pmatrix}}
\newcommand{\eem}{\end{pmatrix}\end{linenomath}}

\newcommand{\minimumEnergy}{\energy{17.8}\xspace}

\newcommand{\firstData}{1st of December 2004\xspace}
\newcommand{\lastData}{31st of December 2012\xspace}

\newcommand{\dEdX}{\ensuremath{\frac{{\rm d}E}{{\rm d}X}}\xspace}

\def\SibyllFull{\textsc{Sibyll2.1}\xspace}
\def\EposFull{\textsc{Epos-LHC}\xspace}
\def\QgIIFull{\textsc{QGSJetII-04}\xspace}

\def\dd{\mathrm{d}}

\graphicspath{{pics/}}

\begin{document}

\title{Depth of Maximum of Air-Shower Profiles at the Pierre Auger Observatory:
   Measurements at Energies above $10^{17.8}$\,\eV}

\author{A.~Aab}
\affiliation{Universit\"{a}t Siegen, Siegen,
Germany}
\author{P.~Abreu}
\affiliation{Laborat\'{o}rio de Instrumenta\c{c}\~{a}o e F\'{\i}sica
Experimental de Part\'{\i}culas - LIP and  Instituto Superior
T\'{e}cnico - IST, Universidade de Lisboa - UL,
Portugal}
\author{M.~Aglietta}
\affiliation{Osservatorio Astrofisico di Torino  (INAF),
Universit\`{a} di Torino and Sezione INFN, Torino,
Italy}
\author{E.J.~Ahn}
\affiliation{Fermilab, Batavia, IL,
USA}
\author{I.~Al Samarai}
\affiliation{Institut de Physique Nucl\'{e}aire d'Orsay (IPNO),
Universit\'{e} Paris 11, CNRS-IN2P3, Orsay,
France}
\author{I.F.M.~Albuquerque}
\affiliation{Universidade de S\~{a}o Paulo, Instituto de F\'{\i}sica,
S\~{a}o Paulo, SP,
Brazil}
\author{I.~Allekotte}
\affiliation{Centro At\'{o}mico Bariloche and Instituto Balseiro
(CNEA-UNCuyo-CONICET), San Carlos de Bariloche,
Argentina}
\author{J.~Allen}
\affiliation{New York University, New York, NY,
USA}
\author{P.~Allison}
\affiliation{Ohio State University, Columbus, OH,
USA}
\author{A.~Almela}
\affiliation{Universidad Tecnol\'{o}gica Nacional - Facultad
Regional Buenos Aires, Buenos Aires,
Argentina}
\affiliation{Instituto de Tecnolog\'{\i}as en Detecci\'{o}n y
Astropart\'{\i}culas (CNEA, CONICET, UNSAM), Buenos Aires,
Argentina}
% ...
\collaboration{The Pierre Auger Collaboration}
\email{auger_spokespersons@fnal.gov}
\noaffiliation

\begin{abstract}
We report a study of the distributions of the depth of
maximum, \Xmax, of extensive air-shower profiles with energies above
\minimumEnergy as observed with the fluorescence telescopes of the
Pierre Auger Observatory. The analysis method for selecting a data
sample with minimal sampling bias is described in detail as well as
the experimental cross-checks and systematic uncertainties. Furthermore,
we discuss the detector acceptance and the resolution of the \Xmax
measurement and provide parameterizations thereof as a function of
energy.  The energy dependence of the mean and standard deviation of
the \Xmax-distributions are compared to air-shower simulations for
different nuclear primaries and interpreted in terms of the mean and
variance of the logarithmic mass distribution at the top of the
atmosphere.
\end{abstract}

\maketitle


\section{Introduction}
\label{sec_intro}

The mass composition of ultra-high energy cosmic rays is one of the
key observables in studies of the origin of these rare particles.
At low and intermediate energies between $10^{17}$ and \energy{19},
precise data on the composition are needed to investigate a potential
transition from galactic to extragalactic sources of the cosmic-ray
flux (see, e.g., \cite{Linsley63, Hillas:2005cs, Allard:2005cx, Aloisio:2006wv}).
Furthermore, the evolution of the composition towards \energy{20} will
help to understand the nature of the steepening of the flux of cosmic
rays observed at around $4{\times}\energy{19}$~\cite{Abbasi:2007sv,
  Abraham:2008ru, AbuZayyad:2012ru}. This flux suppression might
either be caused by energy losses of extragalactic particles due to
interactions with photons (cosmic microwave background in case of
protons or extragalactic background light in case of heavy
nuclei)~\cite{greisen-66, Zatsepin:1966jv}, or it might signify the
maximum of attainable energy in astrophysical accelerators, resulting
in a rigidity-dependent change of the composition (see,
e.g., \cite{peters61, Allard:2008gj, Aloisio:2009sj, Hooper:2009fd,
  Fang:2013cba}).

Due to the low intensity of cosmic rays at the highest energies, the
primary mass cannot be measured directly but is inferred from the
properties of the particle cascade initiated by a primary cosmic ray
interacting with nuclei of the upper atmosphere. These extensive air
showers can be observed with ground-based detectors over large
areas. The mass and energy of the primary particles can be inferred from
the properties of the longitudinal development of the cascade and the
particle densities at the ground after making assumptions about the
characteristics of high-energy interactions (see,
e.g., \cite{Kampert:2012mx} for a recent review).

The energy deposited in the atmosphere by the secondary air-shower
particles is dominated by electron and positron
contributions.  The development of the
corresponding \emph{electromagnetic cascade}~\cite{Greisen:1960wc} is
best described as a function of traversed air mass, usually referred
to as the slant depth $X$.
It is obtained by integrating the density of air along the
direction of arrival of the air shower through the curved atmosphere,
\begin{equation}
  X(z) = \int_z^\infty \!\!\rho(\mathbf{r}(z'))\,\dd z',
\end{equation}
where $\rho(\mathbf{r}(z))$ is the density of air at a point with
longitudinal coordinate $z$ along the shower axis.

The depth at which the energy deposit reaches its maximum is the focus
of this article. The \emph{depth of shower maximum}, \Xmax, is
proportional to the logarithm of the mass $A$ of the primary particle.
However, due to fluctuations of the properties of the first few
hadronic interactions in the cascade, the primary mass
cannot be measured on an event-by-event basis but must be
inferred statistically from the \emph{distribution} of shower maxima
of an ensemble of air showers. Given that nuclei of mass $A_i$ produce
a distribution $f_i(\Xmax)$, the overall \Xmax distribution is
composed of the superposition
\be
  f(\Xmax) = \sum_i p_i \, f_i(\Xmax),
\label{eq:xmaxdistr}
\ee
where the fraction of primary particle of type $i$ is given by $p_i$. The
first two moments of $f(\Xmax)$, i.e., its mean and variance,
\meanXmax and $\sigma(X_\text{max})^2$ respectively, have been extensively studied in
the literature~\cite{elong1, elong2, elong3, LinsleyXmax1983,
  LinsleyXmax1985, semisuper, Matthews:2005sd}. They are to a good
approximation linearly related to the first two moments of the
distribution $g$ of the logarithm of the primary mass $A$, which is given by
the superposition
\be g(\ln A) = \sum_i p_i \, \delta\left(\ln A - \ln A_i\right),
\label{eq:lnadistr}
\ee
where $\delta$ is the Dirac delta function.

The exact shape of $f_i(\Xmax)$ as well as the coefficients
that relate the moments of $g(\ln A)$ to the ones of $f(\Xmax)$ depend on the
details of hadronic interactions in air showers (see,
e.g., \cite{Ulrich:2010rg, Parsons:2011ad, Engel:2011zzb}).  On the one
hand, this introduces considerable uncertainties in the interpretation
of the \Xmax distributions in terms of primary mass, since the
modeling of these interactions relies on extrapolations of accelerator
measurements to cosmic-ray energies. On the other hand, it implies
that the \Xmax distributions can be used to study properties of
hadronic interactions at energies much larger than currently available
in laboratory experiments. A recent example of such a study is the
measurement of the proton-air cross section at $\sqrt{s}=57\,\TeV$
using the upper tail of the \Xmax distribution~\cite{Abreu:2012wt}.

Experimentally, the longitudinal profile of the energy deposit of an
air shower in the atmosphere (and thus \Xmax) can be determined by
observing the fluorescence light emitted by nitrogen molecules excited
by the charged particles of the shower. The amount of fluorescence
light is proportional to the energy deposit~\cite{Belz:2006,
  Ave:2008zza} and can be detected by optical telescopes. The
instrument used in this work is described in Sec.\,\ref{sec_pao} and
the reconstruction algorithms leading to an estimate of \Xmax are laid
out in Sec.\,\ref{sec_reco}.

Previous measurements of \Xmax with the fluorescence technique have
concentrated on the determination of the mean and standard deviation
of the \Xmax distribution~\cite{flysEyeXmax1990, Abbasi:2004nz,
  Abbasi:2009nf, Abraham:2010yv}. Whereas with these two moments the
overall features of primary cosmic-ray composition can be studied, and
composition fractions in a three-component model can even be derived,
only the distribution contains the full information on
composition and hadronic interactions that can be obtained from
measurements of \Xmax.

In this paper, we provide for the first time the measured \Xmax
distributions together with the necessary information to account for
distortions induced by the measurement process.  The relation between
the true and observed \Xmax distribution is
\begin{equation}
\begin{split}
f_\text{obs}&(X_\text{max}^\text{rec}) =\\
&\int_0^\infty \!\!\!
f(\Xmax)\; \varepsilon(\Xmax)\; R(X_\text{max}^\text{rec} - \Xmax)\;
\dd\Xmax,
\end{split}
\label{eq:measurement}
\end{equation}
i.e., the true distribution $f$ is deformed by the detection
efficiency $\varepsilon$ and smeared by the detector resolution $R$
that relates the true \Xmax to the reconstructed one,
$X_\text{max}^\text{rec}$. For an ideal detector, $\varepsilon$ is
constant and $R$ is close to a delta function.  In
Sec.\,\ref{sec_analysis}, we describe the fiducial cuts applied to the
data that guarantee a constant efficiency over a wide range of \Xmax
and the quality cuts that assure a good \Xmax resolution.
Parameterization of $\varepsilon$ and $R$ are given in
Sec.\,\ref{sec_acceptance} and~\ref{sec_resolution}.

Given $f_\text{obs}$, $R$ and $\varepsilon$ it is possible to invert
Eq.~\eqref{eq:measurement} to obtain the true distribution $f(\Xmax)$.
However, since $f_\text{obs}$ is obtained from a limited number of
events, its statistical uncertainties propagate into large
uncertainties and negative correlations of the deconvoluted estimator
of the true distribution, $\hat{f}(\Xmax)$. In practice, methods which
reduce the uncertainties of $\hat{f}(\Xmax)$ by applying additional
constraints to the solution exist~(see, e.g., \cite{Cowan:2002in}), but
these constraints introduce biases that are difficult to quantify.
Therefore we choose to publish $f_\text{obs}$ together with
parameterizations of $R$ and $\varepsilon$. In
Sec.\,\ref{sec_moments} it is demonstrated how to derive \meanXmax and
\sigmaXmax from $f_\text{obs}$, $R$ and $\varepsilon$.  The systematic
uncertainties in the measurement of $f_\text{obs}$ are discussed in
Sec.\,\ref{sec_syst} and validated in Sec.\,\ref{sec_cross}.  In
Sec.\,\ref{sec_results} the measured \Xmax distributions will be shown
in bins of energy reaching from $E=\energy{17.8}$ to
${>}\energy{19.5}$ together with a discussion of their first two
moments.

\section{The Pierre Auger Observatory}
\label{sec_pao}

In this paper we present data from the Pierre Auger
Observatory~\cite{Abraham:2004dt}. It is located in the province of
Mendoza, Argentina, and consists of a Surface-Detector array
(SD)~\cite{Allekotte:2007sf} and a Fluorescence Detector
(FD)~\cite{Abraham:2009pm}. The SD is equipped with over 1600
water-Cherenkov detectors arranged in a triangular grid with a 1500\,m
spacing, detecting photons and charged particles at ground level.
This 3000\,km$^2$ array is overlooked by 24 fluorescence telescopes
grouped in units of 6 at four locations on its periphery. Each
telescope collects the light of air showers over an aperture of
3.8\,m$^2$.  The light is focused on a photomultiplier (PMT) camera with a 13 m$^2$
spherical mirror.  Corrector lenses at the aperture minimize spherical
aberrations of the shower image on the camera.  Each camera is
equipped with 440 PMT pixels, whose field of view is approximately
1.5$^\circ$. One camera covers 30$^\circ$ in azimuth and elevations
range from $1.5^\circ$ to 30$^\circ$ above the horizon. The FD allows
detection of the ultraviolet fluorescence light induced by the energy
deposit of charged particles in the atmosphere and thus measures the
longitudinal development of air showers. Whereas the SD has a duty
cycle near 100\%, the FD operates only during dark nights and under
favorable meteorological conditions leading to a reduced duty cycle of
about 13\%.

Recent enhancements of the Observatory include a sub-array of surface
detector stations with a spacing of 750\,m and three additional
fluorescence telescopes with a field of view from $30^\circ$ to
$60^\circ$ co-located at one of the ``standard'' FD sites~\cite{Amiga,
  HEAT}. These instruments are not used in this work, but they will
allow us to extend the analysis presented here
to lower energies ($E\sim\energy{17}$)~\cite{heatPaper}.

In addition to the FD and SD, important
prerequisites for a precise measurement of the energy and \Xmax of
showers are devices for the calibration of the instruments and the
monitoring of the atmosphere.

The calibration of the fluorescence telescopes in terms of photons at
aperture per ADC count in the PMTs is achieved by approximately yearly
absolute calibrations with a Lambertian light source of known
intensity and nightly relative calibrations with light-emitting diodes
illuminating the FD cameras~\cite{Brack:2004af, Bauleo:2008iz,
  Brack:2013bta}.  The molecular properties of the atmosphere at the
time of data taking are obtained as the 3-hourly data tables provided
by the Global Data Assimilation System (GDAS)~\cite{gdas}, which have
been extensively cross-checked with radio soundings on
site~\cite{Abreu:2012zg}.  The aerosol content of the atmosphere
during data taking is continuously
monitored~\cite{Abraham:2010pf}. For this purpose, the vertical
aerosol optical depth (VAOD) is measured on an hourly basis using
laser shots provided by two central laser facilities
(CLFs)~\cite{Arqueros:2005yn, Abreu:2013qtw} and cross-checked by
lidars~\cite{BenZvi:2006xb} operated at each FD site.  Finally, clouds
are detected via observations in the infrared by cameras installed at
each FD site~\cite{Chirinos:ICRC2013} and data from the Geostationary
Operational Environmental Satellites (GOES)~\cite{goes, Abreu:2013qfa}

\begin{figure*}[t]
\begin{minipage}{0.55\textwidth}
\centering \subfloat[Camera view. The timing of the pixel pulses is
  denoted by shades of gray (early =light, late = dark). The line
  shows the shower detector plane.]
           {\label{fig_cameraPlot}\includegraphics[clip,viewport=-80 -16
               315 235,width=\linewidth]{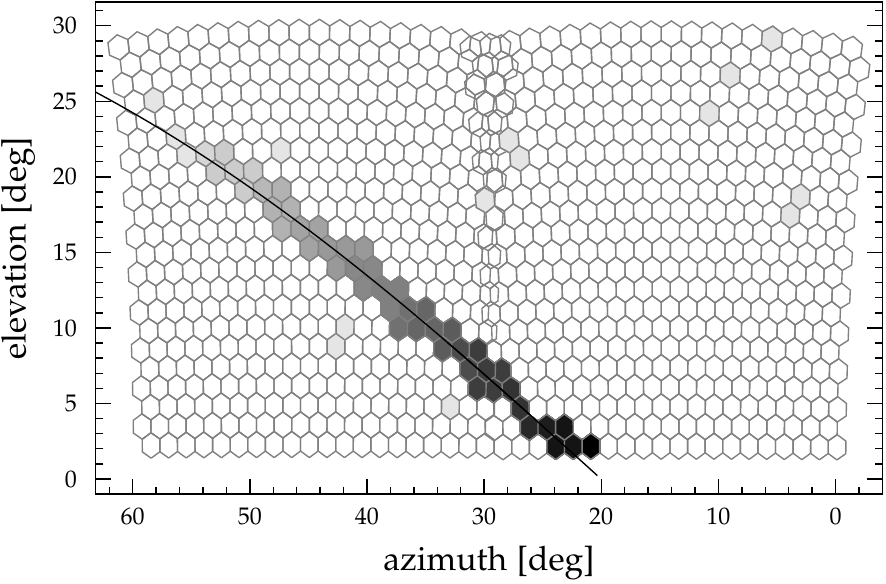}}\\ \subfloat
           [Event Geometry. Pixel viewing angles are shown as shaded
             lines and the shower light and surface detector signals
             are illustrated by markers of different size in
             logarithmic scale.]
           {\label{fig_threeDimPlot}\includegraphics[clip,viewport=0
               -25 289
               135,width=\linewidth]{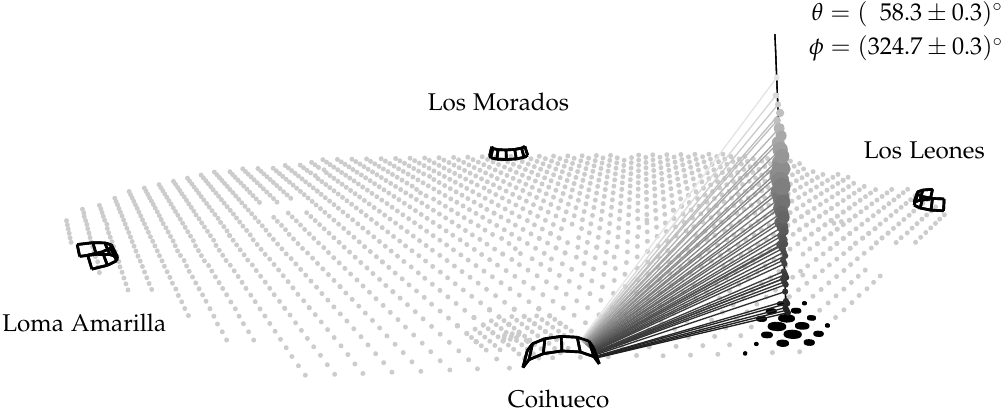}}
\end{minipage}\qquad%
\begin{minipage}{0.4\textwidth}
\centering $\;$\\ \subfloat [Detected photo-electrons (dots) and the
  fitted contributions from components of the shower light (open and
  hatched areas).]
           {\label{fig_lightAtAperture}\includegraphics[clip,viewport=0
               -5 256 193,width=\linewidth]{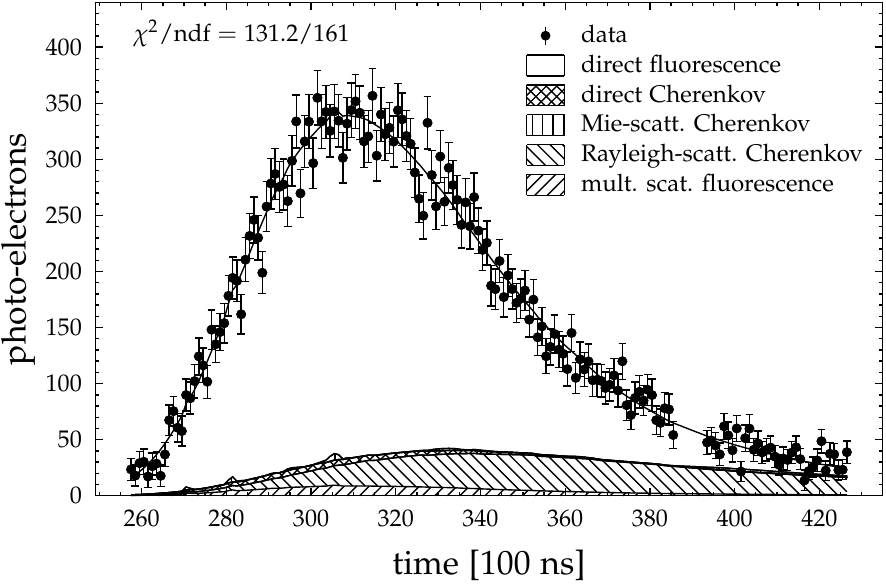}}\\ \subfloat
           [Longitudinal profile (dots) and Gaisser-Hillas function
             (line).]
           {\label{fig_profilePlot}\includegraphics[clip,viewport=0 -5
               256 177,width=\linewidth]{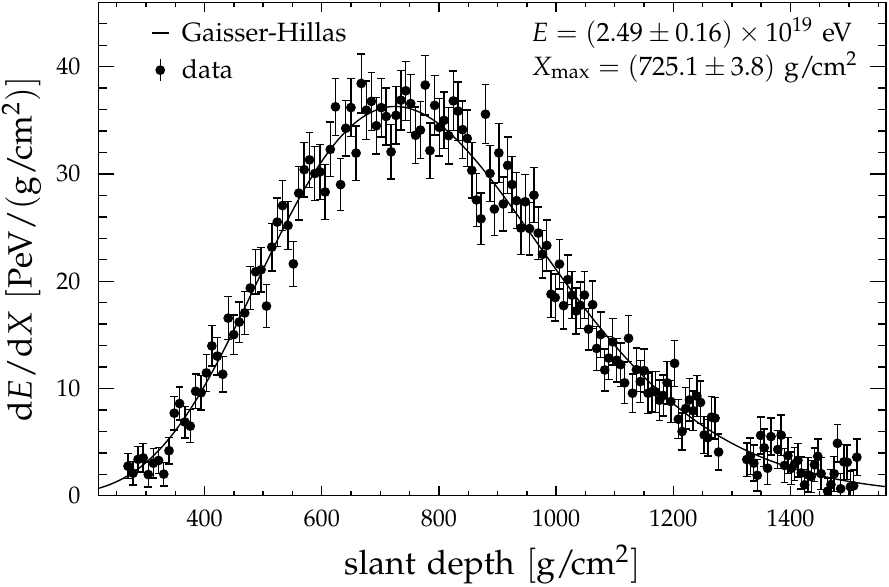}}
  \end{minipage}%
\caption{Reconstruction of event 15346477.}
\label{fig_eventExample}
\end{figure*}

\section{Event Reconstruction}
\label{sec_reco}

The reconstruction of the data is performed within the offline
framework of the Pierre Auger Observatory~\cite{Argiro:2007qg}.
Firstly, all PMT pixels belonging to the shower image on the camera
are identified using a Hough-transformation and subsequently fitted to
reconstruct the plane spanned by the axis of the incoming shower and
the telescope position. Within this plane a three-dimensional
reconstruction of the shower-arrival direction is achieved by
determining the geometry from the arrival times of the shower light as
a function of viewing angle~\cite{Porter:1970et} and from the time of
arrival of the shower front at ground level as measured by the surface-detector station closest to the shower axis. This leads to a
\emph{hybrid} estimate of the shower geometry with a precision of
typically 0.6$^\circ$ for the arrival direction of the primary cosmic
ray~\cite{Sommers:1995dm, Dawson:1996ci, angular}. An example of the
image of a shower in an FD camera is shown in
Fig.\,\ref{fig_cameraPlot} and the reconstructed geometry is shown in
Fig.\,\ref{fig_threeDimPlot}.

The detected signals in the PMTs of the telescope cameras as a
function of time are then converted to a time-trace of light at the
aperture using the calibration of the absolute and relative response
of the optical system. At each time $t_i$, the signals of all PMTs
with pointing directions within an opening angle $\zeta_\text{opt}$
with respect to the corresponding direction towards the shower are
summed up. $\zeta_\text{opt}$ is determined on an event-by-event basis
by maximizing the ratio of the collected signal to the accumulated
noise induced by background light from the night sky. The average
$\zeta_\text{opt}$ of the events used in this analysis is 1.3$^\circ$,
reaching up to 4$^\circ$ for showers detected close to the telescope.
The amount of
light outside of $\zeta_\text{opt}$ due to the finite width of the
shower image~\cite{Gora:2005sq, Giller:2009zz} and the point spread
function of the optical system~\cite{augerPSF1, augerPSF2} is
corrected for in later stages of the reconstruction and
multiply-scattered light within $\zeta_\text{opt}$ is also accounted
for~\cite{Roberts:2005xv, Pekala:2009fe, Giller:2012tt}.

With the help of the reconstructed geometry, every time bin is
projected to a piece of path length $\Delta\ell_i$ on the shower axis
centered at height $h_i$ and slant depth $X_i$. The latter is inferred
by integrating the atmospheric density through a curved atmosphere.
Given the distance to the shower, the light at the aperture can be
projected to the shower axis to estimate the light emitted by the air-shower
particles along $\Delta\ell_i$, taking into account the
attenuation of light due to Rayleigh scattering on air and Mie
scattering on aerosols.

The light from the shower is composed of fluorescence and Cherenkov
photons. The production yield of the former is proportional to the energy
deposited by the shower particles within the volume under study, and
the latter depends on the number of charged particles above the
energy threshold for Cherenkov emission.  Due to the universality of the energy
spectra of electrons and positrons in air showers~\cite{cher:giller,
  cher:hillaslongi,cher:nerling, Lafebre:2009en}, the energy deposit
and the number of particles are proportional, and therefore an exact
solution for the reconstruction of the longitudinal profile of either
of these quantities exists~\cite{Unger:2008uq}.  An example of a
profile of the reconstructed energy deposit can be seen in
Fig.\,\ref{fig_profilePlot} and the contributions of the different
light components to the detected signal are shown in
Fig.\,\ref{fig_lightAtAperture}. The Cherenkov light production is
calculated following~\cite{cher:nerling} and for the fluorescence-light
emission along the shower we use the precise laboratory
measurements of the fluorescence yield from~\cite{Ave:2007xh,
  Ave:2012ifa}.

In the final step of the reconstruction, the shower maximum and total
energy are obtained from a log-likelihood fit of the number of
photo-electrons detected in the PMTs using the Gaisser-Hillas
function~\cite{ghfunc}, $f_\text{GH}$, as a functional description of
the dependence of the energy deposit on slant depth,
\begin{equation}
\begin{split}
f_\text{GH}&(X)=\\
&\left(\dEdX\right)_\text{max}
\left(\frac{X-X_0}{\Xmax-X_0}\right)^\frac{\Xmax-X_0}{\lambda}
\mathrm{e}^{\frac{\Xmax-X}{\lambda}}\,.
\end{split}
  \label{eq:GH}
\end{equation}
The two shape parameters $X_0$ and $\lambda$ are constrained to their
average values to allow for a gradual transition from a two- to a
four-parameter fit depending on the amount of slant depth observed
along the track and the number of detected photons from the respective
event, cf.~\cite{Unger:2008uq}. The constraints are set to the
average values found in the ensemble of events for which an
unconstrained fit with four-parameters is possible. They are given by
$\langle X_0\rangle=-121\,\gcm$ and $\langle\lambda\rangle=61\,\gcm$,
and the observed standard deviations of these sample means are 172 and
13\,\gcm, respectively.  An example of a Gaisser-Hillas function that has been
obtained by the log-likelihood fit to the detected photo-electrons in
Fig.\,\ref{fig_lightAtAperture} is shown in
Fig.\,\ref{fig_profilePlot}.

The calorimetric energy of the shower is obtained by the integration
of $f_\text{GH}$ and the total energy is derived after correcting for
the ``invisible'' energy, carried away by neutrinos and muons. This
correction has been estimated from hybrid data~\cite{eInv} and is of
the order of 10 to 15\% in the energy range relevant for this study.

\section{Data Selection}
\label{sec_analysis}

The analysis presented in this paper is based on data collected by the
Pierre Auger Observatory from the \firstData to the \lastData with the
four standard FD sites.  The initial data set consists of about
$2.6{\times}10^{6}$ shower candidates that met the requirements of
the four-stage trigger system of the data acquisition. Since
only very loose criteria need to be fulfilled at a trigger level
(basically a localized pattern of four pixels detecting a pulse in a
consecutive time order), a further selection of the events is applied
off-line as shown in Tab.~\ref{tab_cutEfficieny} and explained in more
detail in the following section.

\begin{table}[t]
\centering
\caption{Event selection criteria, number of events after each cut and
  selection efficiency with respect to the previous cut.}
\label{tab_cutEfficieny}
\begin{tabular}{lrl}
 cut & \multicolumn{1}{l}{events} & $\varepsilon$
         [\%]\\ \hline
\multicolumn{3}{l}{\emph{pre-selection:}}\\ air-shower candidates &
2573713& -\\ hardware status & 1920584& 74.6\\ aerosols & 1569645&
81.7\\ hybrid geometry & 564324& 35.9\\ profile reconstruction &
539960& 95.6\\ clouds & 432312& 80.1\\ $E>\minimumEnergy$ & 111194&
25.7\\ \hline
\multicolumn{3}{l}{\emph{quality and fiducial
    selection:}}\\ $P(\text{hybrid})$ & 105749& 95.1\\ \Xmax observed
& 73361& 69.4\\ quality cuts & 58305& 79.5\\ fiducial field of view &
21125& 36.2\\ profile cuts & 19947& 94.4
\end{tabular}
\end{table}

\subsection{Pre-Selection}

In the first step, a pre-selection is applied to the air-shower
candidates resulting in a sample with minimum quality requirements
suitable for subsequent physics analysis.

Only time periods with good data-taking conditions are selected using
information from databases and results from off-line quality-assurance
analyses.  Concerning the status of the FD telescopes, a high-quality
calibration of the gains of the PMTs of the FD cameras is required and
runs with an uncertain relative timing with respect to the surface
detector are rejected using information from the electronic logbook
and the slow-control database. Furthermore, data from one telescope
with misaligned optics are not used prior to the date of
realignment. In total, this conservative selection based on the
hardware status removes about 25\% of the initial FD triggers.
Additional database cuts are applied to assure a reliable correction
of the attenuation of shower light due to aerosols: events are only
accepted if a measurement of the aerosol content of the atmosphere is
available within one hour of the time of data taking. Periods with
poor viewing conditions are rejected by requiring that the measured
VAOD, integrated from the ground to
3\,km, is smaller than 0.1. These two requirements reduce the event
sample by 18\%.

For an analysis of the shower maximum as a function of energy, a full
shower reconstruction of the events is needed. The requirement of a
reconstructed hybrid geometry is fulfilled for about 36\% of the
events that survived the cuts on hardware status and atmospheric
conditions. This relatively low efficiency is partially due to
meteorological events like sheet lightning at the horizon that pass
the FD trigger criteria but are later discarded in the event
reconstruction.  Moreover, below $E=\energy{17.5}$
the probability for at least one triggered station in the standard
1.5\,km grid of the surface detector drops quickly~\cite{Abreu:2010aa}. Therefore, a fit
of the geometry using hybrid information is not possible for the
majority of the showers of low energy that trigger the data-acquisition
system of the FD.

\begin{figure*}[t]
\centering \includegraphics[width=0.97\linewidth]{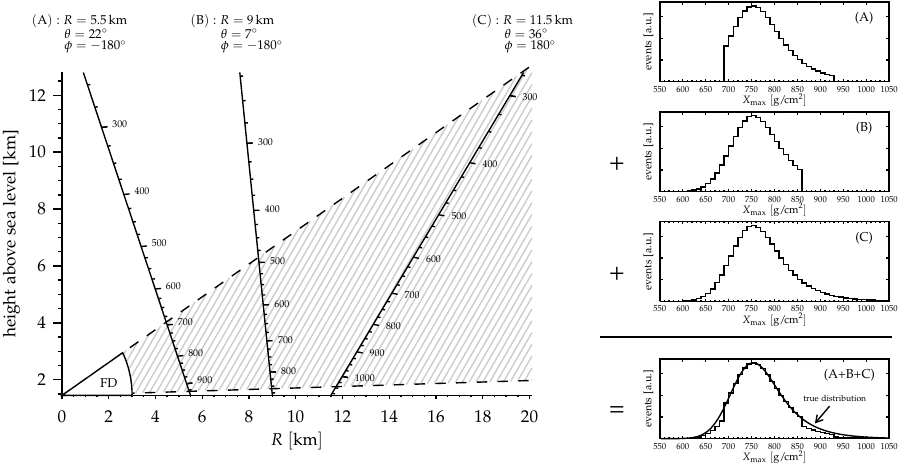}
\caption{Illustration of the influence of the FD field of view on the
  sampling of the \Xmax distribution. The slant depth axes in \gcm are
  shown on the left panel for three different examples of event
  geometries (A), (B) and (C) with different ground distances $R$,
  zenith angle $\theta$ and azimuthal angle $\phi$.  The FD field of
  view is indicated by the hatched area inside the dashed lines.
  Examples of correspondingly truncated \Xmax distributions are shown
  on the right panel together with their sum. For the purpose of this
  illustration, the same number of events for each geometry has been
  assumed.}
\label{fig_fovIllustration}
\end{figure*}

A full reconstruction of the longitudinal profile, including energy
and \Xmax, is possible for most of the events with a hybrid geometry.
Less than 5\% of the events cannot be reconstructed,
because too few profile points are available and/or their
statistical precision is not sufficient. This occurs for either
showers that are far away from the telescopes and close to the trigger
threshold or for geometries pointing towards the telescope for which
the trace of light at the camera is highly time-compressed.

A possible reflection or shadowing of the light from the shower by
clouds is excluded by combining information from the two laser
facilities, the lidars and the cloud monitoring devices described in
Sec.\,\ref{sec_pao}.  Events are accepted if no cloud is detected
along the direction to the shower in either the telescope projection
(cloud camera) or ground-level projection (GOES).  Furthermore, events
are accepted if the base-height of the cloud layer as measured by
both the lidars and lasers is above the geometrical field of view or
400\,\gcm above the fiducial field of view. The latter variable is
explained in the next section. When none of these requirements are
met, events are rejected if either the cloud camera or GOES indicates
the presence of clouds in their respective projections. When no data
from these monitors are available, the event is accepted if during the
data-taking the average cloud fraction as reported by lidars is below
25\%, otherwise the event is not used. In that way, about 80\% of the
events are labeled as cloud-free.

In the final step of the pre-selection, we apply the lower energy
threshold of this analysis, $E>\energy{17.8}$, which reduces the data
set by another 75\% to $1.1{\times}10^5$ events available for further
analysis.

\subsection{Quality and Fiducial Selection}
\label{sec_qAndFidSel}

After the pre-selection described above, the remaining part of the
analysis is focused on defining a subset of the data for which the
distortion of the \Xmax distribution is minimal, i.e., to achieve a
good \Xmax resolution via quality cuts and a uniform acceptance to
showers in a large range of possible \Xmax values.

Before giving the technical details of the selection below, it is
instructive to discuss first some general considerations about the
sampling of the \Xmax distribution with fluorescence detectors.  The
position of the shower maximum can only be determined reliably if the
\Xmax point itself is observed within the field of view of the
telescopes.  The inference of \Xmax from only the rising or falling
edge of the profile would introduce a large dependency of the
results on the functional form of the profile (e.g., Gaisser-Hillas
function) and the constraints on the shape parameters.  The standard
telescopes of the Pierre Auger Observatory are used to observe shower
profiles within elevation angles from 1.5$^\circ$ to 30$^\circ$. This
\emph{geometrical field of view} sets an upper and lower limit on the
range of detectable shower maxima for a given arrival direction and
core location, as illustrated in Fig.\,\ref{fig_fovIllustration} for
three example geometries. Nearby showers with an axis pointing away
from the detector have the smallest acceptance for shallow showers
(geometry (A)), whereas vertical showers cannot be used to sample the
deep tail of the \Xmax distribution for depths larger the vertical depth of the
ground level, which is about 860\,\gcm for the Malarg\"ue site
(geometry (B)).  Ideal conditions for measuring a wide range of \Xmax
are realized by a geometry that intercepts the upper field of view at
low slant depths and by inclined arrival directions, for which the
slant depth at the ground is large (geometry (C)). The true distribution
considered for all three cases is identical and indicated as a solid
line.  If the frequencies of shower maxima detected with all occurring
geometries are collected in one histogram, then the resulting observed
distribution will be under-sampled in the tails at small and large
depths, as illustrated by the (A+B+C)-distribution in the lower right
of Fig.\,\ref{fig_fovIllustration}.

In addition to these simple geometrical constraints, the range of viewable
depths is limited by the following two factors. Firstly,
showers cannot be observed to arbitrary distances, but for a given
energy the maximum viewing distance depends on background light from
the night sky (as a function of elevation) and the transmissivity of
the atmosphere. Therefore, even if shower (C) has a large geometrical
field of view, in general \Xmax will not be detectable at all depths
along the shower axis.  Secondly, if quality cuts are applied to the
data, the available range in depth depends on the selection efficiency
and therefore the corresponding \emph{effective field of view} will
usually be a complicated function of energy, elevation and distance to
the shower maximum.

In this work, we follow a data-driven approach to minimize
the deformation of the \Xmax distributions caused by the effective field
of view boundaries. As will be shown in the
following, a fiducial selection can be applied to the data to
select geometries preferentially with a large accessible field of view
as in the case of the example geometry (C) resulting in an acceptance that
is constant over a wide range of \Xmax values. The different steps of
the quality and fiducial selection are explained in the following.

\paragraph{Hybrid Probability}
After the pre-selection, only events with at least one triggered
station of the SD remain in the data set. The maximum allowed distance
of the nearest station to the reconstructed core is 1.5\,km. For low
energies and large zenith angles, the array is not fully efficient at
these distances. To avoid a possible mass-composition bias due to the
different trigger probabilities for proton- and iron-induced showers,
events are only accepted if the average expected SD trigger
probability is larger than 95\%. The probability is estimated for each event
given its energy, core location and zenith angle (cf.~\cite{Abreu:2011zzd}).
This cut removes about 5\% of events, mainly at low energies.

\paragraph{\Xmax observed}
It is required that the obtained \Xmax is within the observed profile
range. Events where only the rising and/or falling edge of the profile
has been observed are discarded, since in such cases the position of
\Xmax cannot be reliably estimated.  As can be seen in
Tab.~\ref{tab_cutEfficieny}, about 30\% of the events from the tails
of the \Xmax distribution are lost due to the limited field of view of
the FD telescopes.

\begin{figure}[t]
\centering \includegraphics[width=\linewidth]{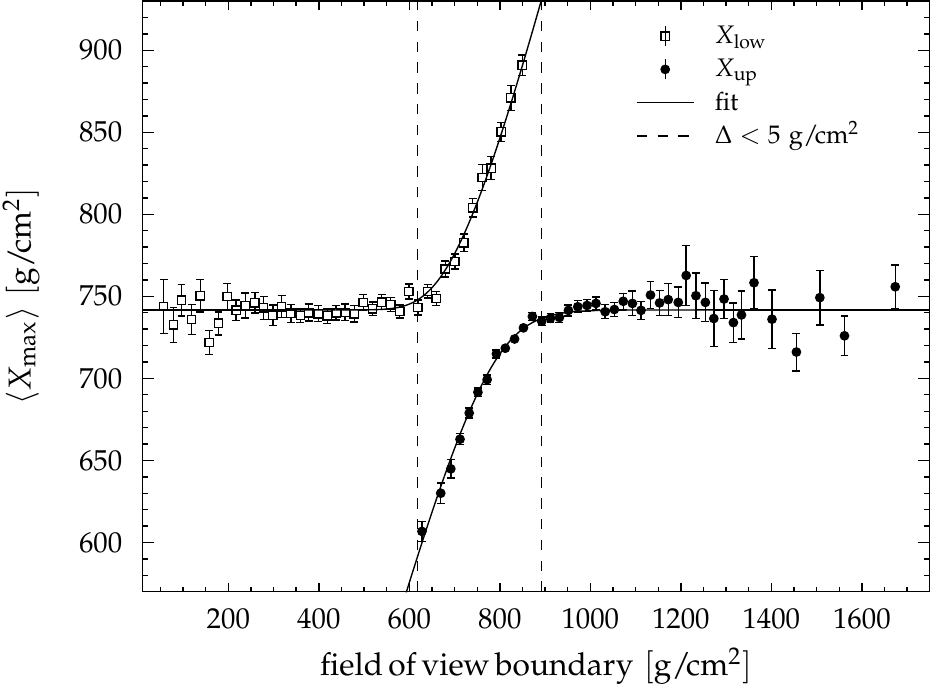}
\caption[fov]{\meanXmax for showers binned in $X_\text{l}$ and
  $X_\text{u}$ in the energy interval $10^{18.1}$ to
  \energy{18.2}. The solid line shows a fit with the truncated mean of
  an exponential function folded with a Gaussian~\cite{Peixoto:2013tu},
  and the dashed line indicates the field-of-view value at which this
  function deviates by more than 5\,\gcm from its asymptotic value.}
\label{fig_fovFit}
\end{figure}

\paragraph{Quality cuts}
Faint showers with a poor \Xmax resolution are rejected based on the
\emph{expected} precision of the \Xmax measurement, $\hat{\sigma}$,
which is calculated in a semi-analytic approach by expanding the
Gaisser-Hillas function around $\Xmax$ and then using this linearized
version to propagate the statistical uncertainties of the number of
photo-electrons at the PMT to an uncertainty of \Xmax. Only showers
with $\hat{\sigma}<40\,\gcm$ are accepted.  Moreover, geometries for
which the shower light is expected to be observed at small angles
with respect to the shower axis are rejected.  Such events exhibit a
large contribution of direct Cherenkov light that falls off
exponentially with the observation angle. Therefore, even small
uncertainties in the event geometry can change the reconstructed
profile by a large amount. We studied the behavior of \meanXmax as a
function of the minimum observation angle, $\alpha_\text{min}$, and
found systematic deviations below $\alpha_\text{min}=20^\circ$, which
is therefore used as a lower limit on the allowed viewing angle.
About 80\% of the events fulfill these quality
criteria.

\paragraph{Fiducial Field of View}
The aim of this selection is to minimize the influence of the
effective field of view on the \Xmax distribution by selecting only
type (C) geometries (cf.\ Fig.\,\ref{fig_fovIllustration}).

The quality variables $\hat{\sigma}$ and $\alpha_\text{min}$ are
calculated for different \Xmax values
in steps of 10\,\gcm along the shower axis within the geometrical
field-of-view boundaries.  In that way, the effective slant-depth
range for high-quality showers can be exactly defined and it is given
by the interval in slant depth for which both
$\hat{\sigma}<40\,\gcm$ and $\alpha_\text{min}>20^\circ$.
The shower
is accepted if this interval is \emph{sufficiently large} to
accommodate the bulk of the \Xmax distribution. The true \Xmax
distribution is unfortunately not known at this stage of the analysis
and therefore we study the differential behavior of \meanXmax on
the lower and upper field-of-view boundary, $X_\text{l}$ and $X_\text{u}$,
for different energy intervals using
data. An example is shown in Fig.\,\ref{fig_fovFit}.  Once the field
of view starts truncating the \Xmax distribution, the observed
\meanXmax deviates from its asymptotically unbiased value.  We set the
          {fiducial field-of-view} boundaries at the values of
          $X_\text{l}$ and $X_\text{u}$ where a deviation of
          $\Delta>5\,\gcm$ occurs to ensure that the overall sampling
          bias on \meanXmax is smaller than this value. The energy
          dependence of these boundaries is then parameterized as
\be
 X_\text{fid}^\text{l,u}(E) =
\begin{cases}
p_1 & ;\lg(E) > p_3,\\ p_1+p_2\,\left(\lg (E) -
p_3\right)^2 & ;\text{otherwise,}
\end{cases}
\label{eq:FOVlow}
\ee
with parameters $\mathbf{p}^\text{u}=(892,-186,18.2)$ and
$\mathbf{p}^\text{l}=(696,$ $-34.6,$ $19.8)$ for the upper and lower
boundary in slant depth, respectively. $p_1$ and $p_2$ are given
in units of \gcm and $E$ is in \eV.  The requirement that $X_\text{l} \leq
X_\text{fid}^\text{l}$ and $X_\text{u} \geq X_\text{fid}^\text{u}$
removes about 64\% of all the remaining events.

\paragraph{Profile Quality}
In the last step of the selection, three more requirements on the
quality of the profiles are applied. Firstly, events with gaps in the
profile that are longer than 20\% of its total observed length are
excluded. Such gaps can occur for showers crossing several
cameras, since the light in each camera is integrated only within the
PMTs that are more than $\zeta_\text{opt}$ away from the camera
border (see, e.g., the gap at around 1300\,\gcm in the profile shown
in Fig.\ref{fig_profilePlot}).  Secondly, residual cloud contamination and horizontal
non-uniformities of the aerosols may cause distortions of the profile
which can be identified with the goodness of the Gaisser-Hillas
fit. We apply a standard-normal transformation to the $\chi^2$ of the
profile fit, $z=(\chi^2-\text{ndf})/\sqrt{2\,\text{ndf}}$, and reject
showers in the non-Gaussian tail at ${>}2.2\,\sigma$. Finally, a
minimum observed track length of ${>}300\,\gcm$ is required.  These
cuts are not taken into account in the calculation of the effective
view, but since the selection efficiency is better than 94\%, the
procedure explained in the last paragraph still yields a good
approximation of the field-of-view boundaries.

In total, the quality and fiducial selection has an efficiency of
18\%. This number is dominated by low-energy showers, where the
profiles are faint and only a small phase space in distance and
arrival direction provides a large effective field of
view. Nevertheless, as shown in Sec.\,\ref{sec_selEff}, the
efficiency of the quality and fiducial selection reaches close to 50\%
at high energies.

\subsection{Final Data Set}

\begin{figure}[!t]
\centering \includegraphics[width=\linewidth]{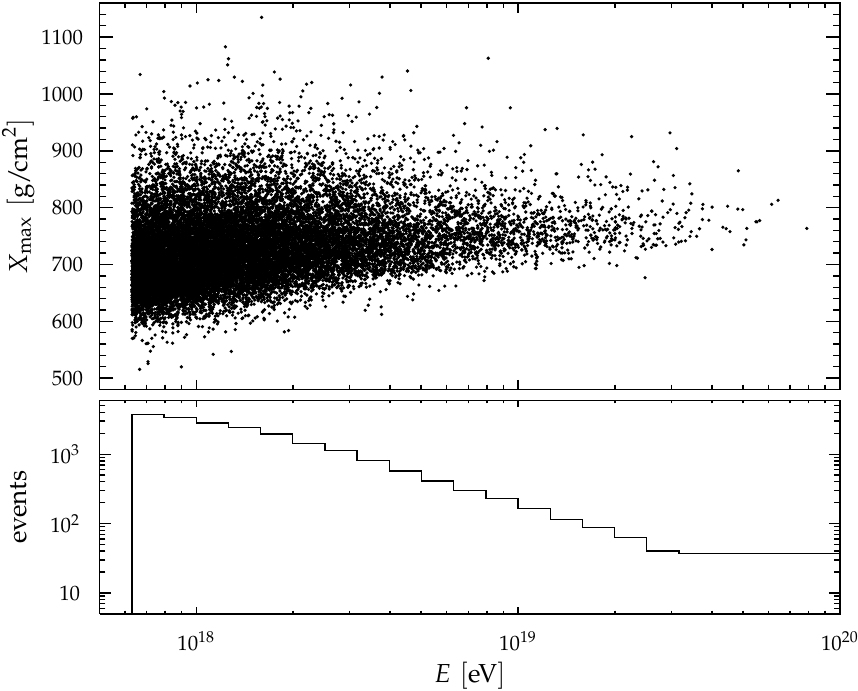}
\caption[scat]{Upper panel: \Xmax and energy of the events used in this paper.
  Lower panel: number of events in bins of energy.}
\label{fig_scatterplot}
\end{figure}

After the application of all selection cuts, 19947 events from the
four standard FD sites remain.  Air showers that have been observed
and selected at more than one FD site are combined by calculating the
weighted average of \Xmax and energy.  This leads to 19759 independent
air-shower events used for this analysis. Their \Xmax and
energy values are shown as a scatter plot in Fig.\,\ref{fig_scatterplot}.

\section{\Xmax Acceptance}
\label{sec_acceptance}

Even following the event selection described above, the probability to
detect and select an air shower is not uniform for
arbitrary values of \Xmax. The corresponding \Xmax acceptance needs to
be evaluated to correct for residual distortions of the \Xmax
distribution. For this purpose we use a detailed, time-dependent
simulation~\cite{Abreu:2010aa} of the atmosphere, the fluorescence and
surface detector. The simulated events are reconstructed with the same
algorithm as the data and the same selection criteria are applied.
The acceptance is calculated from the ratio of selected to generated
events.

The shape of the longitudinal energy-deposit profiles of air showers
at ultra-high energies is, to a good approximation, universal, i.e., it
does not depend on the primary-particle type or details of the first
interaction~\cite{Andringa:2011zz}.  Therefore, after
marginalizing over the distances to the detector and the arrival
directions of the events, the acceptance depends \emph{only} on \Xmax
and the calorimetric energy, but \emph{not} on the primary mass or
hadronic interaction model.  For practical reasons, and since the
calorimetric energies of different primaries with the same total
energy are predicted to be within $\pm 3.5\%$~\cite{Pierog:2013qdx},
we studied the acceptance as a function of total energy and \Xmax.

\begin{figure}[!t]
\centering \includegraphics[width=\linewidth]{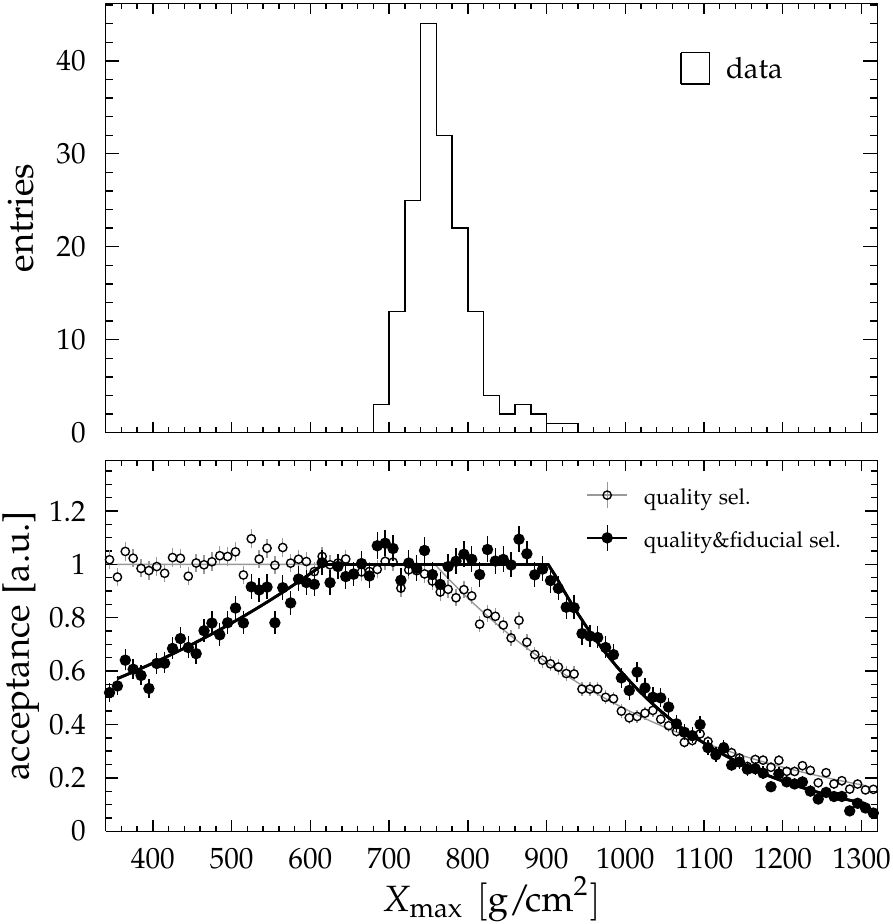}
\caption[acc]{Upper panel: measured \Xmax distribution (full selection,
  $19.0<\lg(E/\!\eV)<19.1$). Lower panel: relative acceptance after
  quality cuts only (open markers) and after quality and fiducial cuts
  (filled markers).
  The parameterizations with Eq.~\eqref{eq:acc} is indicated by lines.}
\label{fig_acceptance}
\end{figure}

In the lower panel of Fig.\,\ref{fig_acceptance} an example of the acceptance
with and
without fiducial field-of-view cuts is shown. Since for the purpose of
the measurement of the \Xmax distribution only the shape of the
acceptance is important, the curves have been normalized to give a
maximum acceptance of 1.  For comparison, the distribution of \Xmax
after the full selection is shown in the upper panel of the figure.  As can be seen, the
acceptance after application of fiducial cuts is constant over most of
the range covered by the selected events. The acceptance without
fiducial selection exhibits a constant part too, but it does not match
the range of measured events well because it starts to depart
from unity already at around the mode of the measured distribution.

Numerically, the \Xmax acceptance can be parameterized by an
exponentially-rising part, a central constant part and an
exponentially-falling part,
\be \varepsilon_\text{rel}(\Xmax) =
  \begin{cases}
    \mathrm{e}^{+\frac{\Xmax-x_1}{\lambda_1}} & ;\Xmax \leq x_1,\\ 1 &
    ;x_1 < \Xmax \leq x_2,\\ \mathrm{e}^{-\frac{\Xmax-x_2}{\lambda_2}}
    & ;\Xmax > x_2,
  \end{cases}
\label{eq:acc}
\ee
with energy-dependent parameters $(x_1, \lambda_1, x_2, \lambda_2)$
that are listed in Tab.~\ref{tab:acceptance}. The uncertainties given
in this table are a combination of statistical and systematic
uncertainties. The former are due to the limited number of simulated
events and the latter are an estimate of the possible changes of the
acceptance due to a mismatch of the optical efficiency, light
production and atmospheric transmission between data and
simulation. The energy scale uncertainty of 14\%~\cite{augerPSF2}
gives an upper limit on the combined influence of these effects and
therefore the systematic uncertainties have been obtained by
re-evaluating the acceptance for simulated events with an energy
shifted by $\pm$14\%.

\section{The Resolution of \Xmax}
\label{sec_resolution}

Besides the acceptance, another important ingredient in the
measurement equation, cf.~Eq.~\eqref{eq:measurement}, is the \Xmax
resolution which determines the broadening of the original
distribution by the statistical fluctuations of $\Xmax^\text{rec}$
around the true \Xmax.  The energy evolution of the \Xmax resolution
is shown in Fig.\,\ref{fig_resolution} where the band denotes its
systematic uncertainty. As can be seen, the total \Xmax resolution is
better than 26\,\gcm at \energy{17.8} and decreases with energy to
reach about 15\,\gcm above \energy{19.3}.  In the following we briefly
discuss the individual contributions to the \Xmax resolution.

\subsection{Detector}
\label{reso_det}

The largest contribution to the \Xmax resolution originates from the
overall performance of the detector system (including the atmosphere) to
collect the light produced by air showers.  The statistical
uncertainty of the determination of the shower maximum from the
Gaisser-Hillas fit, Eq.~\eqref{eq:GH}, is determined by the Poissonian
fluctuations of the number of photo-electrons detected for each
shower. Moreover, the uncertainty of the reconstruction of the arrival
direction of a shower adds another statistical component to the
resolution due to the conversion from the height of the shower maximum
to its slant depth \Xmax. These two contributions can be reliably
determined by a full simulation of the measurement process, including
optical efficiencies and transmission through the
atmosphere~\cite{Prado:2005xj, Abreu:2010aa}. For this purpose we use
showers generated with \textsc{Conex}~\cite{bib:conex} and
\SibyllFull~\cite{Ahn:2009wx} for proton and iron primaries and
re-weight the simulated events to match the observed \Xmax
distribution.  Since high-energy showers are brighter than low-energy
ones, the number of detected photo-electrons increases with energy
and, correspondingly, the resolution improves.  At \energy{17.8}, the
simulations predict a resolution of about 25\,\gcm that decreases to
12\,\gcm towards the highest energies. The systematic uncertainty of
these numbers is of the order of a few~\gcm and has been estimated
by shifting the simulated energies by $\pm$14\% (as previously
explained in the acceptance section).

\begin{figure}[t]
\centering \includegraphics[width=\linewidth]{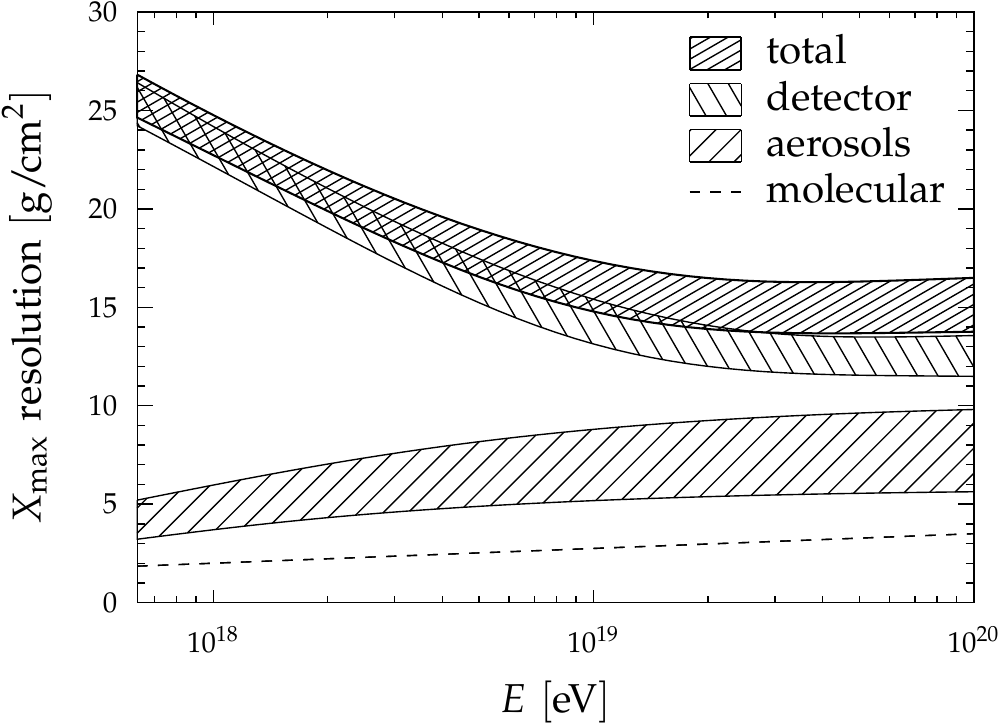}
\caption{\Xmax resolution as a function of energy. Bands denote the
  estimated systematic uncertainties.}
\label{fig_resolution}
\end{figure}

Another detector-related contribution to the resolution originates
from the uncertainties in the alignment of the telescopes.
These are
estimated by comparing the \Xmax values from two reconstructions of
the data set with different alignment constants. One set of constants
has been obtained using the traditional technique of observing tracks
of UV stars (see, e.g., \cite{Sadowski:2002ij}) and the other one used
shower geometries from events reconstructed with the surface detector
for a cross-calibration. The latter are the default constants in the
standard reconstruction.  Averaged over all 24 telescopes, the
$\Delta\Xmax$ values between events from the two reconstructions are found to
be compatible, but systematic alignment differences are present on a
telescope-by-telescope basis giving rise to a standard deviation of
$\Delta\Xmax$ that amounts to $s=\left(5 +
1.1\,\lg(E/\EeV)\right)\,\gcm$. This is used as an estimate of the
contribution of the telescope alignment to the \Xmax resolution by
adding $s/2\pm s/2\, (\text{sys.)}$ to the previously discussed
statistical part of the detector resolution in quadrature.

Finally, uncertainties in the relative timing between the FD
and the SD can introduce additional \Xmax uncertainties, but
even for GPS jitters as large as 100~ns the effect on the \Xmax
resolution is $\lesssim$ 3\,\gcm and can thus be neglected.

The estimated overall contribution of the detector-related
uncertainties to the \Xmax resolution is shown as a back-slashed band
in Fig.\,\ref{fig_resolution}.

\subsection{Aerosols}

Two sources of statistical uncertainty of the aerosol measurements
contribute to the \Xmax resolution. Firstly, the measurement itself is
affected by fluctuations of the night sky background and the number of
photons received from the laser as well as by the time-variability of
the aerosol content within the one-hour averages.  The sum of both
contributions is estimated using the standard deviation of the
quarter-hourly measurements~\cite{Abreu:2013qtw, ValoreICRC13} of the VAOD and
propagated to the \Xmax uncertainty during reconstruction.  Secondly,
non-uniformities of the aerosol layers across the array are estimated
using the differences of the VAOD measurements from different FD sites
and propagated to an \Xmax uncertainty~\cite{Abraham:2010pf}.

The quadratic sum of both sources is shown as the lowest of the dashed
bands in Fig.\,\ref{fig_resolution}, where the systematic uncertainty
given by the width of the band is due to the uncertainty of the
contribution from the horizontal non-uniformity.

\subsection{Molecular Atmosphere}

Finally, the precision to which the density profiles as a function of
height are known gives another contribution to the \Xmax
resolution. It is estimated from the spread of differences between
shower reconstructions using the density profile from GDAS and shower
reconstructions using actual balloon soundings, which are available
for parts of the data (see Fig. 14 in \cite{Abreu:2012zg}).  This
contribution is shown as a dashed line in Fig.\,\ref{fig_resolution}.

\subsection{Parameterization of the Resolution}
The statistical part of the detector resolution arises from the
statistical uncertainty in the determination of \Xmax and from the
statistical uncertainty caused by the conversion from the height of
the maximum in the atmosphere to the corresponding depth of \Xmax.
Simulations of these two contributions show that they are
well-described by the sum of two Gaussian distributions.  The
remaining component to the resolution term of
Eq.~\eqref{eq:measurement} is also Gaussian and describes the
contributions from the calibration of the detector and from the
influence of the atmosphere.  The overall resolution of \Xmax can
therefore be parameterized as
\be
R(X_\text{max}^\text{rec} - \Xmax) = f\;G(\sigma_1) + (1-f)\;
G(\sigma_2)
\label{eq:reso}
\ee
were $G(\sigma)$ denotes a Gaussian distribution with mean zero and
width $\sigma$.  The three parameters $f$, $\sigma_1$ and $\sigma_2$
are listed in Tab.~\ref{tab:resolution} as a function of energy
together with their systematic uncertainties.

\section{\Xmax Moments}
\label{sec_moments}

The parameterized acceptance and resolution together with the measured
\Xmax distributions provide the full information on the shower
development for any type of physics analysis. However, the first two
moments of the distribution, \meanXmax and \sigmaXmax, provide a
compact way to characterize the main features of the distribution.  In
this section we describe three methods that have been explored to
derive the \Xmax moments from our data.

\subsection{Event Weighting}

In this approach each selected shower is weighted according to the
acceptance corresponding to the position of the shower maximum. Events
in the region of constant acceptance are assigned a weight of one. The
under-representation of the distribution in the non-flat part is
compensated for by assigning the inverse of the relative acceptance as
a weight to showers detected in this region, $w = 1 /
\varepsilon_\text{rel}(\Xmax)$, cf.~\eqref{eq:acc}. The unbiased
moments can be reconstructed using the equations for the weighted
moments (cf.~\ref{app_weight}). \sigmaXmax is
then estimated by subtracting the \Xmax resolution in quadrature from
the weighted standard deviation.

\subsection{$\Lambda_\eta$ method}
\label{sec:lambdaeta}
The tail of the \Xmax distribution at large values is related to the
distribution of the first interactions of the primary particles in the
atmosphere (see, e.g., \cite{Ulrich:2009zq}).  Therefore, it is possible
to describe the true distribution of deep showers by an exponential
function.  We subdivide the measured distribution into three regions:
the central part with a constant acceptance, where the distribution
can be measured without distortions, and the shallow and deep regions
where the relative acceptance departs from unity.  Here, for the
purpose of calculating the first two moments of the distribution, the
data are replaced by an exponential function that has been fitted to
the two tails of the distribution, taking the acceptance into account
(see~\ref{app_exppro}). A fraction $\eta$ of the events in the tail is
fitted to obtain the slope $\Lambda_\eta$, similar to the method that
has been used previously to estimate the interaction length of
proton-air collisions~\cite{Baltrusaitis:1984ka, Abreu:2012wt}.  The
mean and standard deviation of the distribution are then calculated by
combining the moments of the undistorted region with the exponential
prolongation in the tails.  In practice, since the \Xmax distributions
have a steep rising edge, the low-\Xmax part is almost fully contained
within the fiducial field of view and only the exponential tail at
deep \Xmax values contributes to a correction with respect to the moments
calculated without taking into account the acceptance.
In the final step, \sigmaXmax is obtained by subtracting
the \Xmax resolution in quadrature from the variance derived with this
procedure.

\subsection{Deconvolution}

As a third method we investigated the possibility to solve
Eq.~\eqref{eq:measurement} for the true \Xmax distribution $f(\Xmax)$
and to subsequently determine the mean and variance of the
solution. For this purpose, Eq.~\eqref{eq:measurement} can be
transformed into a matrix equation by a piece-wise binning in \Xmax
and then be solved by matrix inversion. To overcome the well-known
problem of large variances and negative correlations inherent to this
approach (see, e.g., \cite{Cowan:2002in}), we applied two different
deconvolution algorithms to the data, namely regularized unfolding
using singular value decomposition (SVD) of the migration
matrix~\cite{Hocker:1995kb} and iterative Bayesian
deconvolution~\cite{D'Agostini:1994zf}.

\subsection{Comparison}

Each of these three methods has its own conceptual advantages and
disadvantages.  The main virtue of the event weighting is that it is
purely data-driven. However, with the help of simulated data it was
found that this approach has the largest statistical variance of the
three methods, resulting from large weights that inevitably occur when
a shower is detected in a low-acceptance region.

The estimators of
the moments resulting from the $\Lambda_\eta$-method are also mainly
determined by the measured data since the fiducial field of view
ensures that only the small part of the distribution outside the range
of constant acceptance needs to be extrapolated. The description of
the tail of the distribution with an exponential function has a sound
theoretical motivation. Obviously, this method is not applicable when
the main part of the distribution is affected by distortions from the
acceptance.

Deconvolution is in principle the most mathematically
rigorous method to correct the measured distributions for the
acceptance and resolution.  However, in order to cope with the large
variance of the exact solution, unfolding algorithms need to impose
additional constraints to the data (such as minimal total
curvature~\cite{Tihonov:1963} in case of the SVD approach), that are
less physically motivated than, e.g., an exponential prolongation of
the distribution.

In the following we will use the $\Lambda_\eta$-method as the default
way to estimate the moments of the \Xmax distribution. A comparison
with the results of the other methods will be discussed in
Sec.\,\ref{sec_cross}.  It is worthwhile noting that the moments
calculated without taking into account the acceptance are close to the
ones estimated by the three methods described above, i.e., in the
range of $[0,$ $+3]\,\gcm$ for \meanXmax and $[0,$ $+5]$\,\gcm for
\sigmaXmax.  Assuming a perfect \Xmax resolution would change
$\sigmaXmax$ by $[-5,$ $-3]$\,\gcm.  Thus, the estimates of \meanXmax and
\sigmaXmax are robust with respect to uncertainties on the acceptance
and resolution.

\section{Systematic Uncertainties}
\label{sec_syst}

\subsection{\Xmax Scale}

\begin{figure}[!t]
\centering \includegraphics[width=\linewidth]{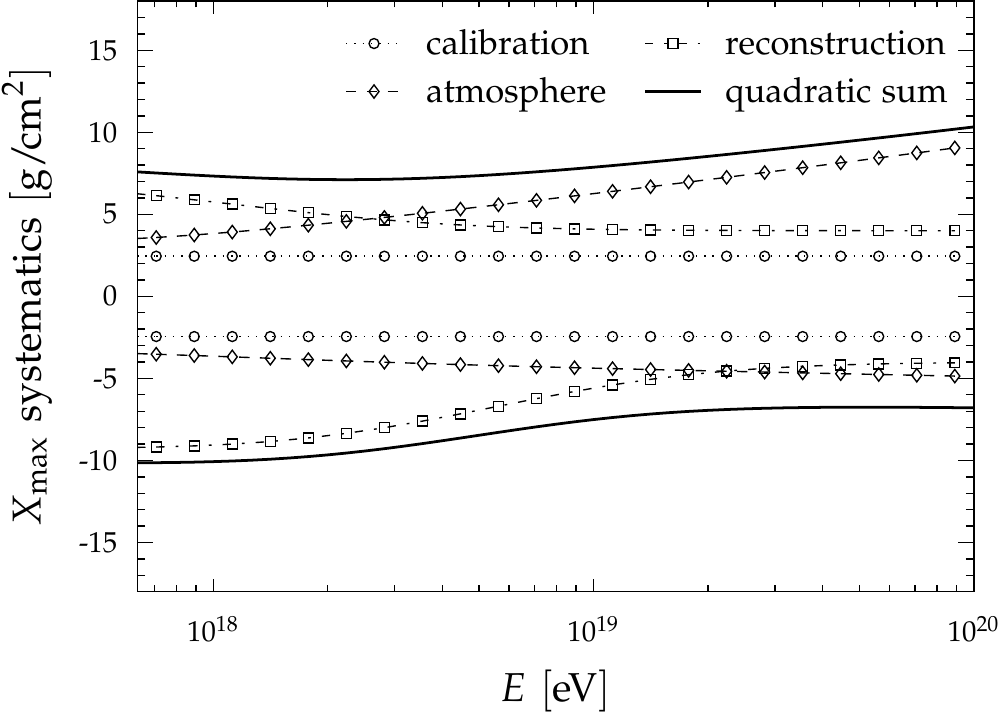}
\caption{Systematic uncertainties in the \Xmax scale as a function of
  energy.}
\label{fig_systematics}
\end{figure}

The systematic uncertainty of the \Xmax scale, i.e., the precision
with which the absolute value of \Xmax can be measured, is shown in
Fig.\,\ref{fig_systematics}.  As can be seen, this uncertainty is
${\leq}10\,\gcm$ at all energies. At low energies, the scale
uncertainty is dominated by the uncertainties in the event
reconstruction and at high energies the atmospheric uncertainties
prevail. The different contributions to the \Xmax scale uncertainty
are discussed in the following. The full covariance
matrix of the \Xmax scale uncertainty is available at~\cite{xmaxDataURL}.

\paragraph{Detector Calibration}
The uncertainties in the relative timing between the FD sites and SD
stations, the optical alignment of the telescopes and the calibration
of the absolute gains of photomultipliers of the cameras
have been found to give only a minor contribution to the \Xmax scale
uncertainty. Their overall contribution is estimated to be less than
3\,\gcm by evaluating the stability of the \Xmax reconstruction under
a variation of the relative timing by its uncertainty of $\pm
100$\,ns~\cite{augerTiming}, using different versions of the gain
calibration and by application of an independent set of alignment
constants (cf.\ Sec.\,\ref{reso_det}).

\paragraph{Reconstruction}
The reconstruction algorithms described in Sec.\,\ref{sec_reco} are
tested by studying the average difference between the reconstructed
and generated \Xmax for simulated data. The \Xmax bias is found to be
less than 3.5\,\gcm and is corrected for during data analysis. The
dependence of the results on the particular choice of function fitted
to the longitudinal profile has been checked by replacing the
Gaisser-Hillas function from Eq.~\eqref{eq:GH} by a Gaussian
distribution in shower age $s=3X/(X+2\Xmax)$, yielding compatible
results within 4\,\gcm for either of the variants proposed
in~\cite{AbuZayyad:2000np} and \cite{gillerGia}. Furthermore, we
tested the influence of the constraints $\langle X_0\rangle$ and
$\langle\lambda\rangle$ used in the Gaisser-Hillas fit by altering
their values by the standard deviations given in Sec.\,\ref{sec_reco},
which changes the \Xmax on average by less than 3.7\,\gcm. Since the
values obtained in these three studies (bias of simulated data, Gaussian
in age and variation of constraints) are just different ways of
assessing the same systematic effect, we do not add them in quadrature
but assign the maximum deviation of 4\,\gcm as an estimate of the
\Xmax scale uncertainty originating from the event reconstruction.

In addition to this validation of the reconstruction of the
longitudinal shower development, we have also studied our
understanding of the lateral distribution of fluorescence and
Cherenkov light and its image on the FD cameras. For this purpose, the
average of the light detected outside the collection angle
$\zeta_\text{opt}$ in data is compared to the amount of light expected
due to the point spread function of the optical system and the lateral
distribution of the light from the shower.  We find that the fraction
of light outside $\zeta_\text{opt}$ is larger in data than in the
expectation and that the ratio of observed-to-expected light depends
on shower age. The corresponding correction of the data during the
reconstruction leads to a shift of \Xmax of $+8.3\,\gcm$ at
\energy{17.8} which decreases to $+1.3\,\gcm$ at the highest
energies. Since the reason for the mismatch between the observed and
expected distribution of the light on the camera is not understood,
the full shift is included as a one-sided systematic uncertainty. With
the help of simulated data we estimated the precision with which the
lateral-light distribution can be measured. This leads to a total
uncertainty from the knowledge of the lateral-light
distribution of $^{+4.7}_{-8.3}\,\gcm$ at \energy{17.8} and
$^{+2.1}_{-1.3}\,\gcm$ at the highest energies.

\paragraph{Atmosphere}
The absolute yield of fluorescence-light production of air showers
in the atmosphere is known with a
precision of 4\%~\cite{Ave:2012ifa}. The corresponding uncertainty
of the relative composition of fluorescence and Cherenkov light leads
to an uncertainty on the shape of the reconstructed longitudinal
profiles and an \Xmax uncertainty of 0.4\,\gcm. Moreover, the
uncertainty in the wavelength dependence of the fluorescence yield
introduces an \Xmax uncertainty of 0.2\,\gcm.
 The amount of
multiply-scattered light to be taken into account during the
reconstruction depends on the shape and size of the aerosols in the
atmosphere. In~\cite{Louedec:2013caa} the systematic effect on the
\Xmax scale has been estimated to be ${\leq}2\,\gcm$.  The systematic
uncertainty of the measurement of the aerosol concentration and its
horizontal uniformity are discussed in~\cite{Abraham:2010pf,
  Abreu:2013qtw, ValoreICRC13}.  They give rise to an energy-dependent systematic
uncertainty of \Xmax, since high-energy showers can be detected at
large distances and have a correspondingly larger correction for the
light transmission between the shower and the detector. Thus, at the
highest energies the \Xmax scale uncertainty is dominated by
uncertainty of the atmospheric monitoring, contributing $^{+7.8}_{-4.2}$\,\gcm in
the last energy bin.

\subsection{\Xmax Moments}
\label{sec_momentsSys}

The systematic uncertainties of \meanXmax and \sigmaXmax are dominated
by the \Xmax scale uncertainty and by the uncertainty of the \Xmax
resolution, respectively, which have been discussed previously
(Sec.\,\ref{sec_syst} and~\ref{sec_resolution}).

In addition, the uncertainties of the parameters of the \Xmax
acceptance, Eq.~\eqref{eq:acc}, are propagated to obtain the
corresponding uncertainties of the moments leading to
${\leq}1.5\,\gcm$ and ${\leq}2.7\,\gcm$ for \meanXmax and \sigmaXmax,
respectively.

Finally, we have also studied the possible bias of the moments
originating from the difference in invisible energy between heavy and
light primaries.  In the energy reconstruction, the \emph{average}
invisible energy is corrected for. If the primary flux is composed of
different nuclei, then the energy of heavy nuclei will be
systematically underestimated and the one of light nuclei will be
overestimated on an event-by-event basis. As a consequence, the
single-nuclei spectra as a function of reconstructed energy will be
shifted with respect to each other and the fraction of nuclei in a bin
of reconstructed energy will be biased.  To study consequences of this
fraction bias on the moments, we consider the extreme case of a
mixture of proton and iron primaries and an invisible energy as
predicted by the \textsc{Epos-LHC} model. The observed energy spectrum
after selection follows, to a good approximation, a power law with a
spectral index $\gamma=-1.76-0.44\,\lg(E/\EeV)$. The potential bias of
the moments due to the invisible energy correction is then found to be
$\delta\meanXmax \leq + 1.2\,\gcm$ and $\delta\sigmaXmax \leq +
0.5\,\gcm$ which we add as a one-sided systematic uncertainty of the
estimated moments.

\begin{figure}[t]
\centering \includegraphics[clip,viewport=0 0 273 224,
  width=\linewidth]{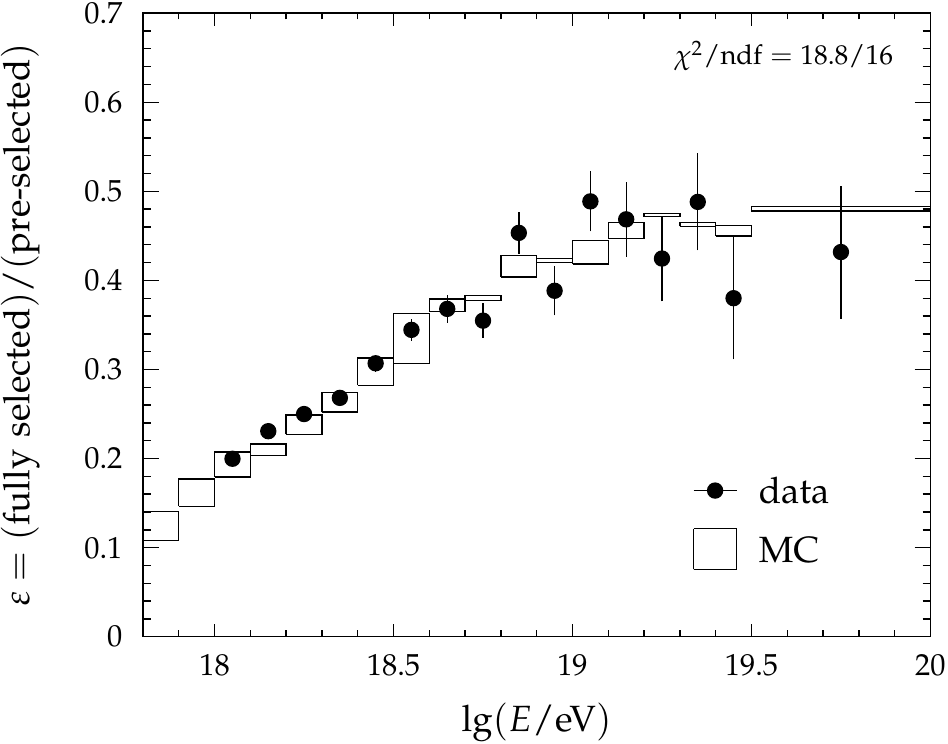}
\caption{Efficiency of the quality and fiducial selection for data and
  MC.  The $\chi^2$ of the sum of the (data-MC) residuals is quoted on
  the top right.}
\label{fig_selEff}
\end{figure}

\section{Cross-Checks}
\label{sec_cross}

The systematic uncertainties estimated in the previous section have
been carefully validated by performing numerous cross-checks on the
stability of the results and the description of the data by the
detector simulation.  In the following we present a few of the most
significant studies.

\subsection{Selection Efficiency}

\label{sec_selEff}
A potential bias from the quality and fiducial selection can be
checked by comparing its efficiency as a function of energy for data
and simulated events.  For this purpose, we use the independent
measurement of air showers provided by the SD and measure the fraction
of events surviving the quality and fiducial cuts out of the total
sample of pre-selected events. This estimate of the selection
efficiency is shown in Fig.\,\ref{fig_selEff} as a function of SD
energy above \energy{18}. Below that energy, the SD trigger efficiency
drops below 50\%. The comparison to the simulated data shows a good
overall agreement and we conclude that the selection efficiency is
fully described by our simulation.

\begin{figure}[t]
\centering \includegraphics[width=0.97\linewidth]{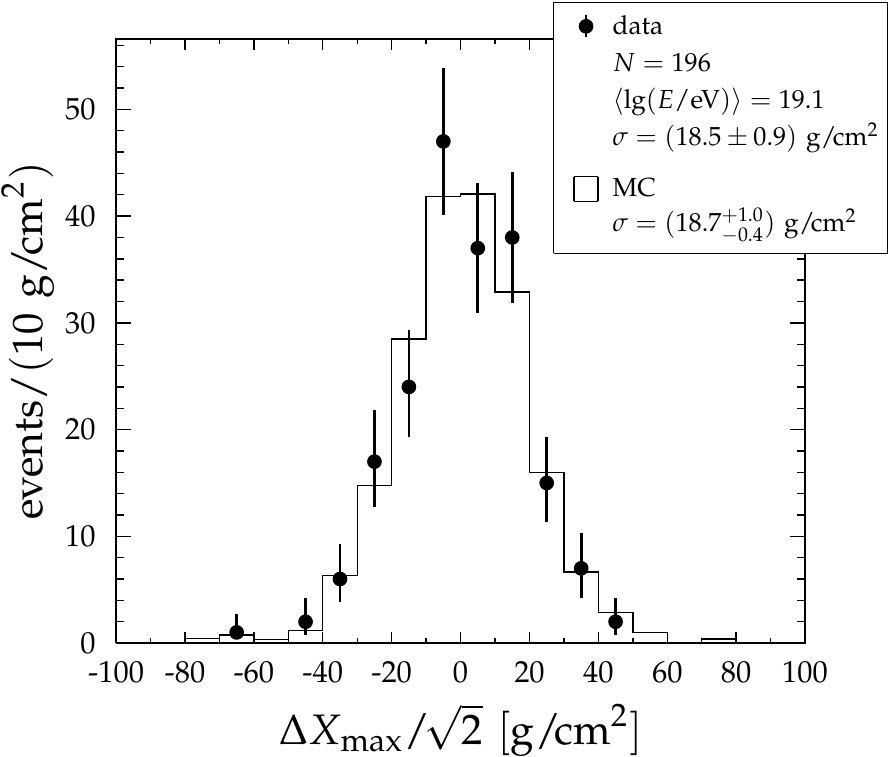}
\caption[stereo]{Distribution of \Xmax differences for events measured by more
  than one FD station. The quoted uncertainties for the standard
  deviation $\sigma$ are statistical for data and systematic for MC. The
  latter are dominated by the uncertainty in the contribution of the alignment
  and aerosols to the resolution (cf.\ Sec.~\ref{sec_resolution}).}
\label{fig_stereo}
\end{figure}

\subsection{Detector Resolution}

The understanding of the detector resolution is checked with the help
of showers that had been detected by more than one FD site. The
distribution of the differences in \Xmax as reconstructed for each
site independently gives an estimate of the \Xmax resolution. As can
be seen in Fig.\,\ref{fig_stereo}, the distribution of the data and
its standard deviation agrees well with the one obtained for simulated
air showers.

\begin{figure*}[!p]
\centering \includegraphics[width=0.8\linewidth]{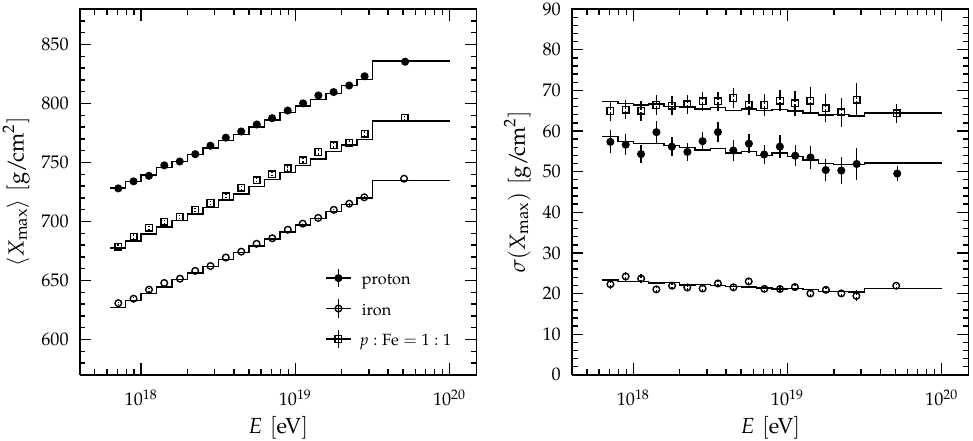}
\caption{Reconstructed \meanXmax and \sigmaXmax (symbols) obtained
  from simulated data for different compositions using the \SibyllFull
  interaction model. The moments of the generated events before
  detector simulation are shown as solid lines.}
\label{fig_mcTest}
\vspace*{0.5cm} \centering
\includegraphics[width=0.9\linewidth]{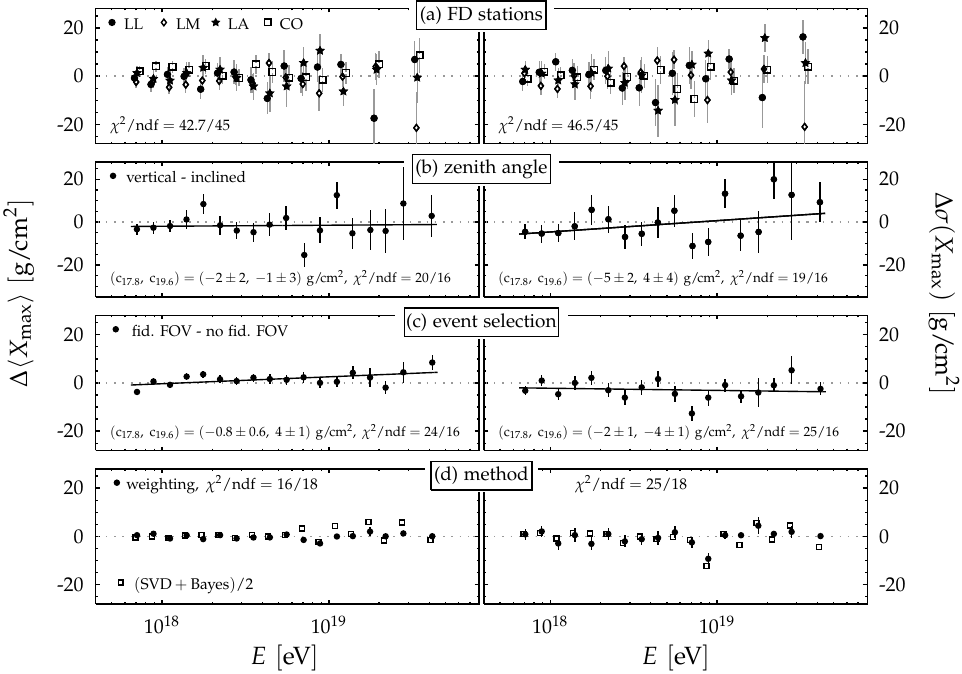}
\caption{Cross-checks. (a) Difference of moments obtained from each FD
  site separately to the results using data from all sites. The global
  $\chi^2$/ndf with respect to zero is given. (b) Subdivision of the
  data set in showers with near-vertical and inclined arrival
  directions. Parameters of a linear fit in $\lg(E)$ are shown with
  supporting points $c_{17.8}$ and $c_{19.6}$ at the centers of the
  first and last bin of $10^{17.85}$ and \energy{19.62}. (c)
  Difference of results with and without fiducial field-of-view
  selection. Parameters are the same as in panel (b). (d) Comparison
  of different methods to estimate \meanXmax and \sigmaXmax. The
  difference from the default method is shown. The average from the two
  deconvolution methods (SVD and Bayesian) is shown without error bars
  (see text). For the weighting method, the $\chi^2$/ndf with respect
  to zero is given. }
\label{fig_crossChecks}
\end{figure*}

\subsection{Analysis of Simulated Data}

The full analysis chain can be validated by applying it to simulated
data and comparing the estimated \Xmax moments to the ones at
generator level. This test has been performed in two variants. In the
first test, we re-evaluated the fiducial field-of-view cuts from the
simulated data to obtain the optimal boundaries with the algorithm
described in Sec.\,\ref{sec_qAndFidSel}. Furthermore, we also tested
the performance when applying the range of the fiducial fields of view
derived from the real data (cf.\ Eq.~\eqref{eq:FOVlow}). This second
test is more conservative as it validates the ability of the analysis
chain to recover the true moments of input distributions it has not
been optimized for. This is an important feature needed for the
comparison of the data to \Xmax distributions that differ from the
observed ones, e.g., for fitting different composition hypotheses to
the data (cf.\ \cite{fractionPaper}).

In both cases, the moments of the input distribution can be reproduced
well. The results from the test using the field-of-view cuts from
Eq.~\eqref{eq:FOVlow} are shown in Fig.\,\ref{fig_mcTest}. As can be
seen, the simulated measurements of \meanXmax and \sigmaXmax agree
within 2\,\gcm with the generated values in case of a pure-proton or
pure-iron composition. Slightly larger biases are visible for a mixed
composition with 50\% proton and 50\% iron where $\meanXmax$ deviates
by about $+4\,\gcm$ from the generated value. This bias can be
partially attributed to the systematic uncertainty of the acceptance
correction and the application of the average invisible-energy correction during
the reconstruction (cf.\ Sec.\,\ref{sec_momentsSys}).  We conclude
that the analysis chain performs well, even for the case where the
cuts of the fiducial fields-of-view are not re-optimized to the input
distributions.

\subsection{FD Sites}

The moments of the \Xmax distribution can be measured for each of the
four FD sites separately to check for possible differences due to
misalignment or systematic differences in the PMT
calibration. Moreover, the four sites (denoted as LL, LM, LA and CO in
the following) are located at different altitudes with a maximum
difference between LL at 1416.2\,m and CO at 1712.3\,m above sea
level. Correspondingly, the aerosols, which have usually their largest
concentration near ground level, are less important for CO than
for the other sites.  The results can be seen in
Fig.\,\ref{fig_crossChecks}~(a), where the differences of the
individual \meanXmax and \sigmaXmax with respect to the results from
the full data sample are shown. A $\chi^2$ test of the compatibility
with zero yields 42.7 and 46.5 for $\Delta$\meanXmax and
$\Delta$\sigmaXmax, respectively. Taking into account that the
comparison is done with the mean of the data, the number of degrees of
freedom is 45 in each case and it can therefore be concluded that the
measurements at the individual sites are indeed statistically-independent
estimates of the same quantity. Averaging the
$\Delta$-values over energy for each station, the maximum deviation
from zero is found to be $2.5\pm1\,\gcm$ for the \meanXmax measured in
CO, which is well within the systematic uncertainties for calibration
and aerosols listed in Sec.\,\ref{sec_syst}.

\subsection{Zenith Angle}

The electromagnetic part of an air shower develops as a function of
traversed air mass. Therefore, the position of the shower maximum
expressed in slant depth does not depend on the zenith angle of the
arrival direction of the cosmic-ray particle.  Accordingly,
\meanXmax and \sigmaXmax are also expected to be independent of the zenith
angle.

However, showers at different zenith angles reach their maximum at
different heights above the ground and in different regions of the
detector acceptance. Therefore, the study of a possible zenith-angle
dependence of the moments of the \Xmax distribution provides an
important end-to-end cross-check of the understanding of the
atmosphere and the detector.

For the purpose of this check, the data set is divided into two
subsamples of approximately equal size at the median zenith angle
$(\cos\theta)_\text{med}=0.795-0.092\,\lg(E/\EeV)$ and the acceptance
and resolution are re-evaluated for these samples.  This yields
estimates of the \Xmax moments for the ``near-vertical'' and
``inclined'' data and their difference is shown in
Fig.\,\ref{fig_crossChecks}~(b). No significant difference is found
over the whole energy range for \meanXmax. At low energies, the
near-vertical \sigmaXmax is smaller by about $5\pm2\,\gcm$ than the
inclined one.  Assuming that either one of the two subsamples gives a
fair estimate of the true width, the corresponding bias of the full
data sample would be $2.2\pm1\,\gcm$, which is compatible with
the systematic uncertainty of the combined \sigmaXmax at low energies.

\subsection{Event Selection}

The dependence of the results on details of the fiducial field of view
as well as on the acceptance and resolution is studied by completely
removing the fiducial field-of-view selection. The data selected in
this way is then corrected with the appropriate acceptance and
resolution using the event weighting method. The difference from the
default moments is shown in Fig.\,\ref{fig_crossChecks}~(c), where the
error bars take into account the correlation between the results due
to the fact that they partially share the same events. As can be seen,
the differences are within 4\,\gcm on average for both, \meanXmax and
\sigmaXmax.  Due to the larger importance of the acceptance correction
in the case of estimating the moments without fiducial cuts, it is
expected that the corresponding systematic uncertainties are larger
than the ones discussed in Sec.\,\ref{sec_syst}. Moreover, the \Xmax
resolution of this selection is worse than the default discussed in
Sec.\,\ref{sec_resolution}.  Given these differences, we conclude that
the two results are in good overall agreement.

\subsection{Analysis Method}

The different methods for the estimation of the \Xmax moments that
were introduced in Sec.\,\ref{sec_moments} are compared in
Fig.\,\ref{fig_crossChecks}~(d). The {event-weighting} yields results
that are very similar to the $\Lambda_\eta$-method. The presented
statistical uncertainties account for the correlation of the two
estimates which use exactly the same data set. The results of the two
methods are found to be compatible with a $\chi^2/\text{ndf}$ of 0.9
and 1.4 for \meanXmax and \sigmaXmax, respectively.

The moments calculated from the deconvoluted \Xmax distributions using
either the Bayesian or SVD method were found to be compatible within
1~\gcm. Therefore, in Fig.\,\ref{fig_crossChecks}~(d) the differences
from the default result are shown for the arithmetic average of the two.
As can be seen, they scatter around zero with no visible systematic
trend.  The statistical uncertainties of these differences have not
been evaluated, but an estimate of their variances can be obtained by
assuming proportionality to the statistical uncertainties
of the default results. A $\chi^2/ \text{ndf}$ of 1 is obtained
when uncertainties are assumed to be 59\% and 90\%
of those given in Tab.~\ref{tab:results} for \meanXmax and \sigmaXmax,
respectively. Therefore, it can be concluded that the moments obtained
by deconvolution agree with the default results within the statistical
uncertainties of the latter.

\begin{figure*}[!p]
\centering \includegraphics[width=\linewidth]{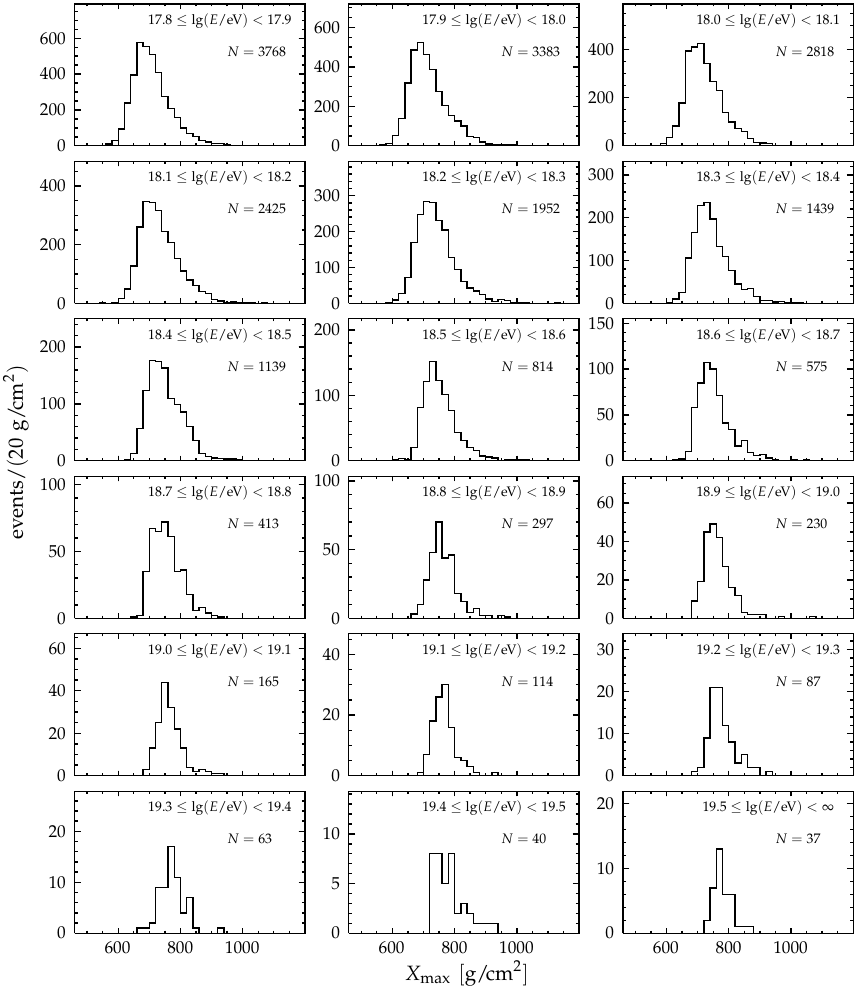}
\caption{\Xmax distributions for different energy intervals.}
\label{fig_distributions}
\end{figure*}

\begin{figure*}[t]
\centering \includegraphics[width=\linewidth]{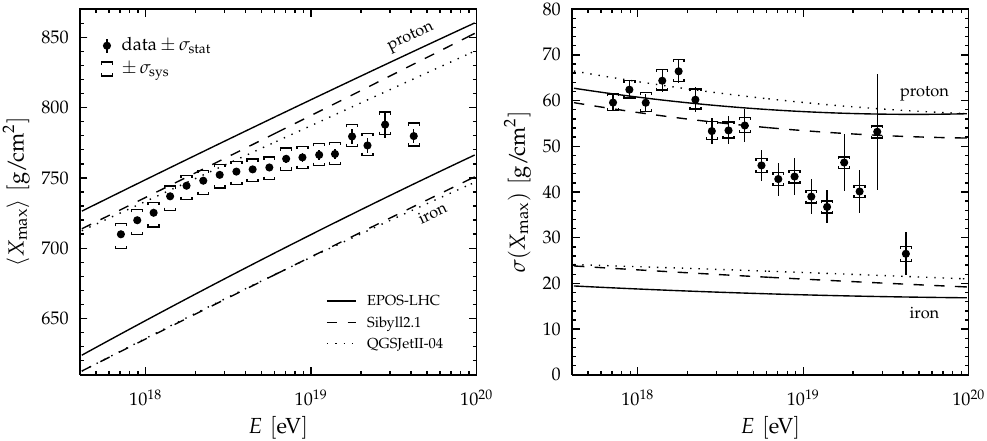}
\caption{Energy evolution of the first two central moments of the
  \Xmax distribution compared to air-shower simulations for proton and
  iron primaries~\cite{bib:conex, Ahn:2009wx, Ostapchenko:2010vb,
    sergeiICRC11, Pierog:2006qv, Pierog:2013ria}.}
\label{fig_moments}
\end{figure*}

\section{Results and Discussion}
\label{sec_results}

In the following we present the results of this analysis in
energy bins of $\Delta\lg(E/\!\eV)=0.1$. Above \energy{19.5} an
integral bin is used.  The highest-energy event in this data sample
had been detected by all four FD sites and its reconstructed energy
and shower maximum are $E=(7.9\pm0.3){\times}10^{19}$\,\eV\ and
$\Xmax=762\pm2\,\gcm$, respectively, where the uncertainties are statistical only.

The \Xmax distributions after event selection are shown in
Fig.\,\ref{fig_distributions}. These are the ``raw'' distributions
($f_\text{obs}(X_\text{max}^\text{rec})$ in
Eq.~\eqref{eq:measurement}) that still include effects of the detector
resolution and the acceptance.  Electronically readable tables of the
distributions, as well as the parameters of the resolution and
acceptance, are available at~\cite{xmaxDataURL}.  A thorough
discussion of the distributions can be found in an accompanying
paper~\cite{fractionPaper}, where a fit of the data with simulated
templates for different primary masses is presented.

In this paper we will concentrate on the discussion of the first two
moments of the \Xmax distribution, \meanXmax and \sigmaXmax, which are
listed in Tab.~\ref{tab:results} together with their statistical and
systematic uncertainties.  The statistical uncertainties are
calculated with the parametric bootstrap method. For this purpose, the
data are fitted with Eq.~\ref{eq:measurement} assuming the functional
form suggested in \cite{Peixoto:2013tu} as $f(\Xmax)$. Given this
parametric model of the true \Xmax distribution, realizations of the
measurement are repeatedly drawn from Eq.~\ref{eq:measurement} with
the number of events being equal to the ones observed.  After
application of the $\Lambda_\eta$ analysis described in
Sec.~\ref{sec:lambdaeta}, distributions of \Xmax and \sigmaXmax are
obtained from which the statistical uncertainties of the measured
moments are estimated.

A comparison of the predictions of the moments from
simulations for proton- and iron-induced air showers to the data is shown in
Fig.\,\ref{fig_moments}. The simulations have been performed using the
three contemporary hadronic interaction models that were either tuned
to recent LHC data (\QgIIFull~\cite{Ostapchenko:2010vb, sergeiICRC11},
\EposFull~\cite{Pierog:2006qv, Pierog:2013ria}) or found in good
agreement with these measurements (\SibyllFull\, \cite{Ahn:2009wx}, see
\cite{d'Enterria:2011kw}). It is worth noting that the energy of
the first data point in Fig.\,\ref{fig_moments} corresponds to a
center-of-mass energy that is only four times larger than the one
currently available at the LHC ($\sqrt{s}=8\,\TeV$). Therefore, unless
the models have deficiencies in phase-space regions that are not
covered well by LHC measurements, the uncertainties due to the
extrapolation of hadronic interactions to the lower energy threshold
of this analysis should be small.  On the other hand, the last energy
bin at $\langle\lg(E/\!\eV)\rangle=19.62$ corresponds to a center-of-mass
energy that is a factor of about 40 higher than the LHC energies and
the model predictions have to be treated more carefully.

Comparing the energy evolution of \meanXmax for data and simulations
in Fig.\,\ref{fig_moments} it can be seen that the slope of the data
is different than what would be expected for either a pure-proton or
pure-iron composition. The change of \meanXmax with the logarithm of
energy is usually referred to as \emph{elongation rate}~\cite{elong1,
  elong2, elong3},
\be D_{10} = \frac{\dd\meanXmax}{\dd\lg (E/\!\eV)}.  \ee
Within the superposition model, where it is assumed that a primary
nucleus of mass $A$ and energy $E$ can be to a good approximation
treated as a superposition of $A$ nucleons of energy $E^\prime = E/A$,
the elongation rate is expected to be the same for any type of primary.
Any deviation of an observed elongation rate from this expectation
$\hat{D}_{10}$ can be attributed to a change of the primary
composition,
\be
 D_{10} = \hat{D}_{10} \left(1-\frac{\dd\meanLnA}{\dd\ln(E/\!\eV)}\right).
\ee
\begin{figure*}[t]
\centering \includegraphics[width=\linewidth]{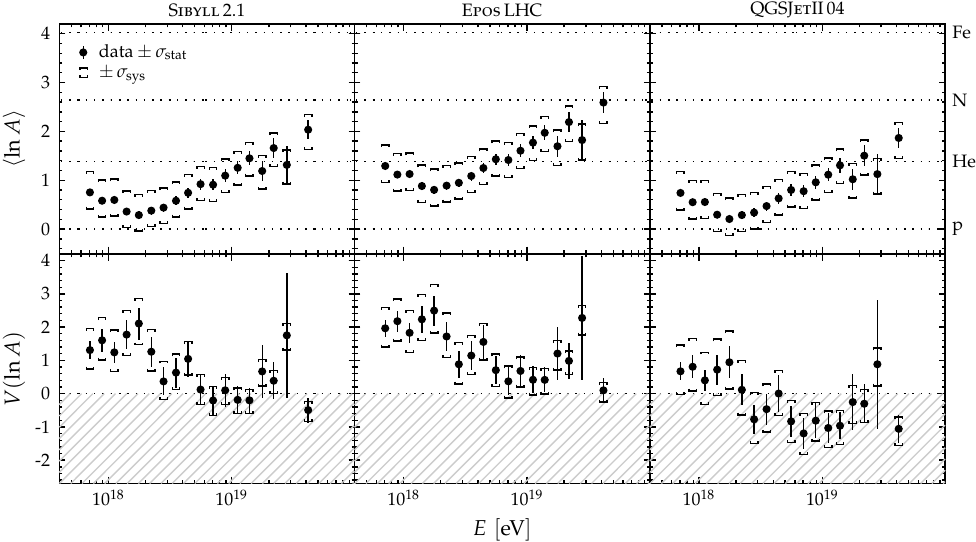}
\caption{Average of the logarithmic mass and its variance estimated
  from data using different interaction models. The non-physical
  region of negative variance is indicated as the gray dashed region.}
\label{fig_lnA}
\end{figure*}

A single linear fit of \meanXmax as a function of $\lg(E)$ does not
describe our data well ($\chi^{2}/\text{ndf}=138.4/16$). Allowing for
a change in the elongation rate at a break point $\lg(E_0)$ yields a
good $\chi^{2}/\text{ndf}$ of $8.2/14$ with an elongation rate of
\be D_{10}=86.4\pm
5.0\,\text{(stat.)}\,^{+3.8}_{-3.2}\,\text{(sys.)}\,\gcmdec
\label{D10_below}
\ee
below $\lg(E_0/\!\eV)=18.27\pm
0.04\,\text{(stat.)}\,^{+0.06}_{-0.07}\,\text{(sys.)}$ and
\be D_{10}=26.4\pm
2.5\,\text{(stat.)}\,^{+7.0}_{-1.9}\,\text{(sys.)}\,\gcmdec
\label{D10_above}
\ee
above this energy. The average shower maximum at $E_0$ is $746.8\pm
2.1\,\text{(stat.)}\,^{+\phantom{1}6.6}_{-10.0}\,\text{(sys.)}$\,\gcm. Here
the systematic uncertainties on $D_{10}$ have been obtained by varying
the individual contributions of the systematic uncertainties on
\meanXmax separately.

The elongation rates predicted by air-shower simulations for a
constant composition range from 54 to 64\,\gcmdec. Together with the
results in Eqs.~\eqref{D10_below} and \eqref{D10_above} we can
therefore deduce that
\be
\frac{\dd\meanLnA}{\dd\lg (E/\!\eV)} = -1.07 \pm
0.20\,\text{(stat.)}\,^{+0.15}_{-0.13}\,\text{(sys.)}\,^{+0.26}_{-0.31}\,\text{(model)}
\ee
below $E_0$ and
\be
\frac{\dd\meanLnA}{\dd\lg (E/\!\eV)} = +1.23 \pm
0.10\,\text{(stat.)}\,^{+0.07}_{-0.27}\,\text{(sys.)}\,^{+0.09}_{-0.10}\,\text{(model)}
\ee
above this energy.  This implies that there is an evolution of the
average composition of cosmic rays towards lighter nuclei up to
energies of \energy{18.27}.  Above this energy, the trend reverses and
the composition becomes heavier.

A similar behavior is visible for the width of the \Xmax distribution
in the right panel of Fig.\,\ref{fig_moments}, where it can be seen
that the \sigmaXmax gets narrower towards high energies, as it would
be expected for showers induced by heavy nuclei.

For a more quantitative study of the evolution of the composition,
\meanXmax and \sigmaXmax are converted to the first two moments of the
$\ln A$ distribution (cf.\ Eq.~\eqref{eq:lnadistr}) following the
method described in~\cite{Abreu:2013env, AhnICRC13}. The mean and variance of
$\ln A$ are shown in Fig.\,\ref{fig_lnA} using air-shower simulations
with three interaction models. As can be seen for all three cases, the
composition is lightest at around \energy{18.3} and the different
features of hadronic interactions implemented in the three models give
rise to differences in \meanLnA of about $\pm 0.3$. The interpretation
with \EposFull leads to the heaviest average composition that is
compatible with the $\ln A$ of nitrogen at the highest energies.  The
variance of $\ln A$ derived with \EposFull and \SibyllFull suggests
that the flux of cosmic rays is composed of different nuclei at low
energies and that it is dominated by a single type of nucleus above
\energy{18.7} where the variance, $V(\ln A)$, is close to zero. The
interpretation with \QgIIFull leads to unphysical variances ($V(\ln
A)<0$) above \energy{18.4} and therefore this model is disfavored by
our data, unless one allows for a systematic bias that is twice as
large as the uncertainties estimated in Sec.\,\ref{sec_syst}.\\
\section{Conclusions}

In this paper, we presented the measurement of the distribution of the
depth of shower maximum of ultra-high energy cosmic-ray air
showers. We described the data selection which allows for a nearly
unbiased measurement of the distributions and discussed the residual
effects of acceptance and resolution. The data set is the largest
sample of \Xmax measurements hitherto collected by a cosmic-ray
detector. We provide computer-readable tables of the distributions and
detector parameters that make it possible to interpret the
measurements without the need of additional software to simulate the
detector response. This approach will also facilitate the comparison
with measurements of \Xmax from other
experiments~\cite{Barcikowski:2013nfa}.  Here we cannot provide such a
comparison, since for these data neither the detector bias is
controlled for using fiducial cuts, nor are the resolution and
acceptance publicly available.

An interpretation in terms of mass composition of the moments of the
\Xmax distribution was given using air-shower simulations with
contemporary hadronic interaction models. Assuming that the modeling
of hadronic interactions gives a fair representation of the actual
processes in air showers at ultra-high energies, our data suggest that
the flux of cosmic rays is composed of predominantly light nuclei at
around \energy{18.3} and that the fraction of heavy nuclei is
increasing up to energies of \energy{19.6}.  Estimates of the
fractions of groups of nuclei contributing to the cosmic-ray flux can
be derived by interpreting the full distributions.
Such an analysis can be found in an accompanying paper~\cite{fractionPaper}.

\section{Acknowledgments}

The successful installation, commissioning, and operation of the Pierre Auger Observatory would not have been possible without the strong commitment and effort from the technical and administrative staff in Malarg\"{u}e.

We are very grateful to the following agencies and organizations for financial support:
Comisi\'{o}n Nacional de Energ\'{\i}a At\'{o}mica, Fundaci\'{o}n Antorchas, Gobierno De La Provincia de Mendoza, Municipalidad de Malarg\"{u}e, NDM Holdings and Valle Las Le\~{n}as, in gratitude for their continuing cooperation over land access, Argentina; the Australian Research Council; Conselho Nacional de Desenvolvimento Cient\'{\i}fico e Tecnol\'{o}gico (CNPq), Financiadora de Estudos e Projetos (FINEP), Funda\c{c}\~{a}o de Amparo \`{a} Pesquisa do Estado de Rio de Janeiro (FAPERJ), S\~{a}o Paulo Research Foundation (FAPESP) Grants \# 2010/07359-6, \# 1999/05404-3, Minist\'{e}rio de Ci\^{e}ncia e Tecnologia (MCT), Brazil; MSMT-CR LG13007, 7AMB14AR005, CZ.1.05/2.1.00/03.0058 and the Czech Science Foundation grant 14-17501S, Czech Republic;  Centre de Calcul IN2P3/CNRS, Centre National de la Recherche Scientifique (CNRS), Conseil R\'{e}gional Ile-de-France, D\'{e}partement Physique Nucl\'{e}aire et Corpusculaire (PNC-IN2P3/CNRS), D\'{e}partement Sciences de l'Univers (SDU-INSU/CNRS), Institut Lagrange de Paris, ILP LABEX ANR-10-LABX-63, within the Investissements d'Avenir Programme  ANR-11-IDEX-0004-02, France; Bundesministerium f\"{u}r Bildung und Forschung (BMBF), Deutsche Forschungsgemeinschaft (DFG), Finanzministerium Baden-W\"{u}rttemberg, Helmholtz-Gemeinschaft Deutscher Forschungszentren (HGF), Ministerium f\"{u}r Wissenschaft und Forschung, Nordrhein Westfalen, Ministerium f\"{u}r Wissenschaft, Forschung und Kunst, Baden-W\"{u}rttemberg, Germany; Istituto Nazionale di Fisica Nucleare (INFN), Ministero dell'Istruzione, dell'Universit\`{a} e della Ricerca (MIUR), Gran Sasso Center for Astroparticle Physics (CFA), CETEMPS Center of Excellence, Italy; Consejo Nacional de Ciencia y Tecnolog\'{\i}a (CONACYT), Mexico; Ministerie van Onderwijs, Cultuur en Wetenschap, Nederlandse Organisatie voor Wetenschappelijk Onderzoek (NWO), Stichting voor Fundamenteel Onderzoek der Materie (FOM), Netherlands; National Centre for Research and Development, Grant Nos.ERA-NET-ASPERA/01/11 and ERA-NET-ASPERA/02/11, National Science Centre, Grant Nos. 2013/08/M/ST9/00322, 2013/08/M/ST9/00728 and HARMONIA 5 - 2013/10/M/ST9/00062, Poland; Portuguese national funds and FEDER funds within COMPETE - Programa Operacional Factores de Competitividade through Funda\c{c}\~{a}o para a Ci\^{e}ncia e a Tecnologia, Portugal; Romanian Authority for Scientific Research ANCS, CNDI-UEFISCDI partnership projects nr.20/2012 and nr.194/2012, project nr.1/ASPERA2/2012 ERA-NET, PN-II-RU-PD-2011-3-0145-17, and PN-II-RU-PD-2011-3-0062, the Minister of National  Education, Programme for research - Space Technology and Advanced Research - STAR, project number 83/2013, Romania; Slovenian Research Agency, Slovenia; Comunidad de Madrid, FEDER funds, Ministerio de Educaci\'{o}n y Ciencia, Xunta de Galicia, European Community 7th Framework Program, Grant No. FP7-PEOPLE-2012-IEF-328826, Spain; Science and Technology Facilities Council, United Kingdom; Department of Energy, Contract No. DE-AC02-07CH11359, DE-FR02-04ER41300, DE-FG02-99ER41107 and DE-SC0011689, National Science Foundation, Grant No. 0450696, The Grainger Foundation, USA; NAFOSTED, Vietnam; Marie Curie-IRSES/EPLANET, European Particle Physics Latin American Network, European Union 7th Framework Program, Grant No. PIRSES-2009-GA-246806; and UNESCO.\\

\bibliographystyle{elsarticle-num}
\bibliography{xmax}

\appendix

\section{Calculation of \Xmax Moments}

\subsection{Weighted Events}
\label{app_weight}

One possibility to correct for the acceptance as a function of \Xmax
is to assign to each event a weight $w_i = 1 /
\varepsilon_\text{rel}(X_{\text{max},i})$ . The average shower maximum
of events weighted by the inverse of the acceptance is given by
\be \meanXmax = \textstyle\left(\sum\limits_{i} w_i
\;X_{\text{max},i}\right) \bigg/ \sum_i w_i.  \ee
The second non-central moment is
\be \langle \Xmax^2 \rangle = \textstyle\left(\sum\limits_{i} w_i
\;(X_{\text{max},i})^2\right) \bigg/ \sum_i w_i \ee
with which
\be \sigmaXmax^2 = \textstyle k\, \left(\langle \Xmax^2 \rangle -
\meanXmax^2\right) \ee
where
\be k = \textstyle \left(\textstyle\sum_i w_i\right)^2 \big/
\left(\left(\textstyle\sum_i w_i\right)^2 - \textstyle\sum_i
w_i^2\right) \ee
giving us the usual factor of $k=N/(N-1)$ when all weights are equal
to one.

\begin{figure}[t]
\centering \includegraphics[width=\linewidth]{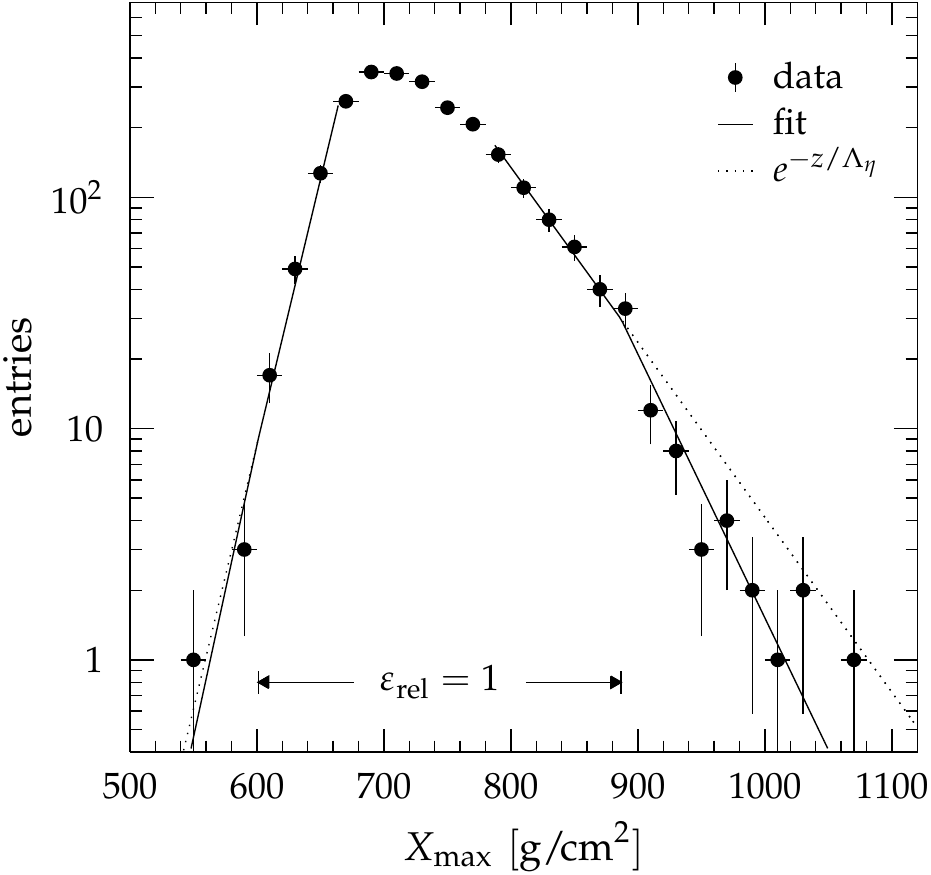}
\caption{Fit to the tails of the \Xmax distribution
  ($18.1<\lg(E/\!\eV)<18.2$). The region of constant acceptance
  $\varepsilon_\text{rel}=1$ is indicated by arrows.}
\label{fig_momentsIllustration}
\end{figure}

\subsection{$\Lambda_\eta$-Method}
\label{app_exppro}

When the shower maxima of the events in the tails of the \Xmax
distribution follow an exponential distribution, damped by an
exponential acceptance above a certain depth
(cf.\ Eq.~\eqref{eq:acc}), then the resulting distribution of the
upper tail is given by
\be f(z) = k\,\mathrm{e}^{-\frac{z}{\Lambda_\eta}}
\begin{cases}
  1 & ;z < z_0,\\ \displaystyle\mathrm{e}^{-\frac{z-z_{0}}{\lambda}} &
  ;\text{otherwise,}
\end{cases}
\ee
and a similar formula describes the lower tail, where $z$ denotes the
distance to the start point of the fit and $z_0$ is the distance above
which the acceptance decreases exponentially with decay constant
$\lambda$.  The normalization is given by
\be k = \Lambda_\eta\, \left(1+\mathrm{e}^{-\frac{z_0}{\Lambda_\eta}}
\left[\frac{\lambda}{\lambda+\Lambda_\eta} -1\right]\right).  \ee
The fraction of events in the tail is denoted by $\eta$.  Following
\cite{Abreu:2012wt} we use $\eta=0.20$ for the tail at large \Xmax and
the leading edge of the \Xmax distribution is fitted using
$\eta=0.15$.

The unbinned likelihood for $N$ events in the tail is
\be -\log\mathcal{L} \propto N\,\log k(\Lambda_\eta) +
\frac1\Lambda_\eta \sum_{i=1}^N z_i, \ee
where terms independent of $\Lambda_\eta$ have been omitted.

An illustration of a fit of the upper and lower tail of the \Xmax
distribution is shown in Fig.\,\ref{fig_momentsIllustration}. The
fitted damped exponential is shown as the solid line and the range of
constant acceptance is indicated by arrows. For the purpose of
calculating the moments, the data distribution is replaced by the
exponential functions (shown as dashed lines) outside of the
$\varepsilon_\text{rel}=1$ range.

\section{Data tables}
\begin{table}[h!]
    \centering
  \caption[result]{Parameters of $\varepsilon_\text{rel}(\Xmax)$ (Eq.~\eqref{eq:acc}) in \gcm.}
  \label{tab:acceptance}
  \begin{tabular}{lllll}
    $\lg E$ range & $x_1$ & $\lambda_1$ & $x_2$ & $\lambda_2$
    \\ \hline $[17.8,17.9)$&         586 $\pm$ \phantom{1}6&         109 $\pm$ \phantom{1}17&         881 $\pm$ \phantom{1}8& \phantom{1}95 $\pm$           7\\
$[17.9,18.0)$&         592 $\pm$ \phantom{1}9&         133 $\pm$ \phantom{1}17&         883 $\pm$ \phantom{1}8&            101 $\pm$           7\\
$[18.0,18.1)$&         597 $\pm$            11&         158 $\pm$ \phantom{1}19&         885 $\pm$ \phantom{1}8&            107 $\pm$           7\\
$[18.1,18.2)$&         601 $\pm$            14&         182 $\pm$ \phantom{1}21&         887 $\pm$ \phantom{1}8&            113 $\pm$           7\\
$[18.2,18.3)$&         604 $\pm$            17&         206 $\pm$ \phantom{1}24&         888 $\pm$ \phantom{1}8&            119 $\pm$           7\\
$[18.3,18.4)$&         605 $\pm$            20&         230 $\pm$ \phantom{1}28&         890 $\pm$ \phantom{1}8&            125 $\pm$           7\\
$[18.4,18.5)$&         605 $\pm$            23&         253 $\pm$ \phantom{1}32&         892 $\pm$ \phantom{1}8&            131 $\pm$           7\\
$[18.5,18.6)$&         604 $\pm$            27&         276 $\pm$ \phantom{1}38&         894 $\pm$ \phantom{1}9&            137 $\pm$           8\\
$[18.6,18.7)$&         602 $\pm$            30&         299 $\pm$ \phantom{1}44&         896 $\pm$ \phantom{1}9&            143 $\pm$           8\\
$[18.7,18.8)$&         599 $\pm$            33&         321 $\pm$ \phantom{1}51&         898 $\pm$ \phantom{1}9&            150 $\pm$           8\\
$[18.8,18.9)$&         594 $\pm$            36&         344 $\pm$ \phantom{1}59&         899 $\pm$ \phantom{1}9&            156 $\pm$           8\\
$[18.9,19.0)$&         588 $\pm$            39&         365 $\pm$ \phantom{1}67&         901 $\pm$ \phantom{1}9&            162 $\pm$           8\\
$[19.0,19.1)$&         581 $\pm$            43&         386 $\pm$ \phantom{1}77&         903 $\pm$ \phantom{1}9&            168 $\pm$           8\\
$[19.1,19.2)$&         573 $\pm$            46&         407 $\pm$ \phantom{1}86&         905 $\pm$ \phantom{1}9&            174 $\pm$           8\\
$[19.2,19.3)$&         563 $\pm$            49&         428 $\pm$ \phantom{1}98&         907 $\pm$ \phantom{1}9&            180 $\pm$           8\\
$[19.3,19.4)$&         553 $\pm$            52&         447 $\pm$            109&         908 $\pm$ \phantom{1}9&            186 $\pm$           8\\
$[19.4,19.5)$&         540 $\pm$            56&         468 $\pm$            122&         910 $\pm$ \phantom{1}9&            192 $\pm$           8\\
$[19.5,\infty)$&         517 $\pm$            62&         502 $\pm$            146&         913 $\pm$            10&            203 $\pm$           9\\

  \end{tabular}
\end{table}
\begin{table}[h!]
   \centering
  \caption[result]{Parameters of the \Xmax resolution (Eq.~\eqref{eq:reso}).
    $\sigma_{1}$ and $\sigma_2$ are in \gcm. The uncertainties are systematic
   and fully correlated between  $\sigma_1$ and $\sigma_2$.}
  \label{tab:resolution}
  \begin{tabular}{llll}
   $\lg E$ range & $\sigma_1$ & $\sigma_2$ & $f$\\ \hline
    $[17.8,17.9)$&        17.5 $\pm$         0.7&        33.7 $\pm$         1.4&        0.62\\
$[17.9,18.0)$&        16.7 $\pm$         0.7&        32.9 $\pm$         1.4&        0.63\\
$[18.0,18.1)$&        15.9 $\pm$         0.7&        31.9 $\pm$         1.4&        0.63\\
$[18.1,18.2)$&        15.1 $\pm$         0.7&        31.0 $\pm$         1.4&        0.64\\
$[18.2,18.3)$&        14.4 $\pm$         0.7&        30.0 $\pm$         1.4&        0.65\\
$[18.3,18.4)$&        13.8 $\pm$         0.7&        29.1 $\pm$         1.5&        0.66\\
$[18.4,18.5)$&        13.3 $\pm$         0.7&        28.1 $\pm$         1.6&        0.67\\
$[18.5,18.6)$&        12.8 $\pm$         0.8&        27.1 $\pm$         1.6&        0.68\\
$[18.6,18.7)$&        12.3 $\pm$         0.8&        26.3 $\pm$         1.7&        0.69\\
$[18.7,18.8)$&        12.0 $\pm$         0.8&        25.4 $\pm$         1.8&        0.70\\
$[18.8,18.9)$&        11.7 $\pm$         0.9&        24.7 $\pm$         1.9&        0.70\\
$[18.9,19.0)$&        11.5 $\pm$         0.9&        24.1 $\pm$         1.9&        0.71\\
$[19.0,19.1)$&        11.3 $\pm$         0.9&        23.6 $\pm$         1.9&        0.72\\
$[19.1,19.2)$&        11.2 $\pm$         0.9&        23.3 $\pm$         2.0&        0.73\\
$[19.2,19.3)$&        11.1 $\pm$         0.9&        23.1 $\pm$         2.0&        0.74\\
$[19.3,19.4)$&        11.1 $\pm$         1.0&        23.1 $\pm$         2.0&        0.75\\
$[19.4,19.5)$&        11.1 $\pm$         1.0&        23.2 $\pm$         2.0&        0.76\\
$[19.5,\infty)$&        11.2 $\pm$         1.0&        23.7 $\pm$         2.1&        0.77\\

  \end{tabular}
\end{table}
\begin{table}[h!]
    \centering
  \caption[result]{First two moments of the \Xmax distributions.
    Energies are in $[\!\eV]$ and \meanXmax and \sigmaXmax are given
    in $[\gcm]$ followed by their statistical and systematic
    uncertainties. The number of selected events in each energy bin is
    given in the third column.}
  \label{tab:results}
  \begin{tabular}{llrll}
     $\lg E$ range & $\langle\lg E\rangle$ &
    \multicolumn{1}{l}{$N$} & \meanXmax & \sigmaXmax\\ \hline
    $[17.8,17.9)$& 17.85&  3768&  709.9 $\pm\phantom{1}1.2  ^{\phantom{1}+7.6}_{      -10.2}$ &   59.6 $\pm\phantom{1}1.7  ^{\phantom{1}+1.9}_{\phantom{1}-1.7}$\\
$[17.9,18.0)$& 17.95&  3383&  719.9 $\pm\phantom{1}1.4  ^{\phantom{1}+7.5}_{      -10.2}$ &   62.4 $\pm\phantom{1}2.1  ^{\phantom{1}+2.1}_{\phantom{1}-1.8}$\\
$[18.0,18.1)$& 18.05&  2818&  725.2 $\pm\phantom{1}1.5  ^{\phantom{1}+7.4}_{      -10.2}$ &   59.5 $\pm\phantom{1}2.0  ^{\phantom{1}+2.2}_{\phantom{1}-1.9}$\\
$[18.1,18.2)$& 18.15&  2425&  736.9 $\pm\phantom{1}1.8  ^{\phantom{1}+7.3}_{      -10.1}$ &   64.3 $\pm\phantom{1}2.6  ^{\phantom{1}+2.4}_{\phantom{1}-2.1}$\\
$[18.2,18.3)$& 18.25&  1952&  744.5 $\pm\phantom{1}2.0  ^{\phantom{1}+7.3}_{\phantom{1}-9.9}$ &   66.4 $\pm\phantom{1}2.6  ^{\phantom{1}+2.6}_{\phantom{1}-2.2}$\\
$[18.3,18.4)$& 18.35&  1439&  748.0 $\pm\phantom{1}2.0  ^{\phantom{1}+7.3}_{\phantom{1}-9.7}$ &   60.2 $\pm\phantom{1}2.8  ^{\phantom{1}+2.3}_{\phantom{1}-2.0}$\\
$[18.4,18.5)$& 18.45&  1139&  752.2 $\pm\phantom{1}2.1  ^{\phantom{1}+7.3}_{\phantom{1}-9.4}$ &   53.3 $\pm\phantom{1}2.9  ^{\phantom{1}+2.1}_{\phantom{1}-1.8}$\\
$[18.5,18.6)$& 18.55&   814&  754.5 $\pm\phantom{1}2.2  ^{\phantom{1}+7.3}_{\phantom{1}-9.1}$ &   53.5 $\pm\phantom{1}3.0  ^{\phantom{1}+1.9}_{\phantom{1}-1.7}$\\
$[18.6,18.7)$& 18.65&   575&  756.1 $\pm\phantom{1}2.7  ^{\phantom{1}+7.4}_{\phantom{1}-8.8}$ &   54.5 $\pm\phantom{1}3.5  ^{\phantom{1}+1.7}_{\phantom{1}-1.6}$\\
$[18.7,18.8)$& 18.75&   413&  757.4 $\pm\phantom{1}2.8  ^{\phantom{1}+7.5}_{\phantom{1}-8.5}$ &   45.8 $\pm\phantom{1}3.4  ^{\phantom{1}+1.5}_{\phantom{1}-1.5}$\\
$[18.8,18.9)$& 18.85&   297&  763.6 $\pm\phantom{1}2.9  ^{\phantom{1}+7.7}_{\phantom{1}-8.1}$ &   42.8 $\pm\phantom{1}3.6  ^{\phantom{1}+1.4}_{\phantom{1}-1.4}$\\
$[18.9,19.0)$& 18.95&   230&  764.6 $\pm\phantom{1}3.2  ^{\phantom{1}+7.8}_{\phantom{1}-7.8}$ &   43.4 $\pm\phantom{1}4.1  ^{\phantom{1}+1.3}_{\phantom{1}-1.4}$\\
$[19.0,19.1)$& 19.05&   165&  766.4 $\pm\phantom{1}3.3  ^{\phantom{1}+8.0}_{\phantom{1}-7.6}$ &   39.0 $\pm\phantom{1}3.8  ^{\phantom{1}+1.3}_{\phantom{1}-1.4}$\\
$[19.1,19.2)$& 19.14&   114&  767.0 $\pm\phantom{1}3.6  ^{\phantom{1}+8.2}_{\phantom{1}-7.4}$ &   36.7 $\pm\phantom{1}3.6  ^{\phantom{1}+1.3}_{\phantom{1}-1.4}$\\
$[19.2,19.3)$& 19.25&    87&  779.5 $\pm\phantom{1}5.1  ^{\phantom{1}+8.5}_{\phantom{1}-7.2}$ &   46.4 $\pm\phantom{1}6.2  ^{\phantom{1}+1.2}_{\phantom{1}-1.3}$\\
$[19.3,19.4)$& 19.34&    63&  773.1 $\pm\phantom{1}5.0  ^{\phantom{1}+8.7}_{\phantom{1}-7.1}$ &   40.1 $\pm\phantom{1}4.8  ^{\phantom{1}+1.3}_{\phantom{1}-1.4}$\\
$[19.4,19.5)$& 19.45&    40&  787.9 $\pm\phantom{1}9.6  ^{\phantom{1}+8.9}_{\phantom{1}-7.0}$ &   53.2 $\pm     12.7  ^{\phantom{1}+1.3}_{\phantom{1}-1.4}$\\
$[19.5,\;\,\infty\;\,)$& 19.62&    37&  779.8 $\pm\phantom{1}5.0  ^{\phantom{1}+9.4}_{\phantom{1}-6.9}$ &   26.5 $\pm\phantom{1}4.8  ^{\phantom{1}+1.5}_{\phantom{1}-1.6}$\\

  \end{tabular}
\end{table}

\end{document}